\documentclass[11pt]{article}

\usepackage{amsfonts}
\usepackage{amssymb}
\usepackage{amsmath}
\usepackage{authblk}
\usepackage{hyperref}
\usepackage{graphicx}
\usepackage{float}
\usepackage{overpic}
\usepackage{geometry}
\usepackage{epstopdf}
\usepackage{multirow}
\usepackage{subcaption}
\usepackage[official]{eurosym}
\providecommand{\U}[1]{\protect\rule{.1in}{.1in}}
\usepackage{caption}
\usepackage{subcaption}
\usepackage{verbatim}

\newtheorem{theorem}{Theorem}[section]

\newtheorem{lemma}[theorem]{Lemma}

\newtheorem{proposition}[theorem]{Proposition}
\newtheorem{remark}[theorem]{Remark}

\newtheorem{assumption}[theorem]{Assumption}

\newenvironment{proof}[1][Proof]{\noindent\textbf{#1.} }{\ \rule{0.5em}{0.5em}}
\usepackage{amsmath,systeme}
\usepackage{lineno}
\usepackage[dvipsnames]{xcolor}
\usepackage{indentfirst}

\DeclareMathOperator*{\argmax}{arg\,max}

\definecolor{p}{rgb}{0.50,0.55,1.00}
\definecolor{tiziano}{rgb}{0.55,0.71,0.0}
\definecolor{alm}{rgb}{0.64,0.06,0.70}
\definecolor{b}{rgb}{0,0,0}

\title{Optimal Investment
and Fair Sharing Rules for 
 Incentives\\
in Virtual Renewable Energy Communities}
\author{Almendra Awerkin\footnote{Department of Mathematics ``Tullio Levi Civita”, University of Padova, {\tt awerkin@math.unipd.it}}  \and Paolo Falbo\footnote{Department of Economics and Management, University of Brescia, {\tt paolo.falbo@unibs.it}} \and Tiziano Vargiolu\footnote{Department of Mathematics ``Tullio Levi Civita”, University of Padova, {\tt vargiolu@math.unipd.it}\\
\noindent  This work was initiated while the first author had a 1-year post-doc position at the University of Brescia from February 2022 to February 2023. 
The second author is supported by the Italian research funding PRIN 2022, project 2022MNK5JZ "Compensation Rules Among Members of Heterogeneous Energy Communities".
The third author 
is supported by the INdAM - GNAMPA Project code CUP E53C23001670001 and by the projects funded by the European Union - Next Generation EU, Mission 4 Component 1, 2022BEMMLZ ``Stochastic control and games and the role of information'' CUP C53D23002430006 and P20224TM7Z ”Probabilistic methods for energy transition”,  CUP C53D23008390001. 
The authors wish to thank for fruitful discussions 
Roberto Baviera, 
Marta Castellini, 
Katia Colaneri, 
Carme Frau, 
Barbara Guardabascio, 
Claudia Nunes, 
Cristian Pelizzari.} }

\begin{document}
\maketitle

\begin{abstract}

This paper examines two interconnected challenges in Renewable Energy Communities (REC) optimization: investment in renewable technologies and equitable sharing of incentives, offered (by a central authority) under the Virtual Framework of self-consumption. Focusing on a REC composed of a household and a biogas producer—common in rural and urban contexts—we analyze how investment decisions and incentive-sharing mechanisms impact community profitability. The household invests in photovoltaic panels to reduce energy purchases and monetize surplus generation, while the biogas producer either converts biogas into electricity or sells it on the gas market. 

We model this interaction as a leader-follower problem: an administrator (leader) defines the incentive-sharing rule, while an household and a biogas (followers) determine their optimal investments. Modeling the objective of the leader as a Nash bargaining problem and a Nash equilibrium for the followers' static game, we provide insights into optimal REC design and policy implications for fostering sustainable community energy systems.

The model is applied to the case of a REC, under realistic data about the random variables and investment costs. We obtain deep insight of key relevance to address both the organization of a REC, the optimal investment decisions for the members and the incentive policy design for the central authority. 
\end{abstract}

\section{Introduction} 

The global challenge of climate change and the ongoing energy transition represent an unprecedented issue for humanity. Addressing this challenge requires diverse solutions that engage all sectors of society. One promising approach to reducing CO$_2$ emissions is empowering end-users to become both self-producers and self-consumers of electricity generated from Renewable Energy Sources (RES). 

In this context, the European Union (EU) has introduced two key directives—Directive (EU) 2018/2001 \cite{2018/2001} and Directive (EU) 2019/944 \cite{2019/944}—which establish the regulatory framework enabling private households, small enterprises, and local public authorities to form Renewable Energy Communities (RECs). These communities facilitate the shared generation and consumption of renewable energy.
The potential impact of RECs is substantial: in OECD European countries, approximately 450 million people collectively consumed over 3 trillion kWh of electricity in 2020 \cite{EIA}. 

Heterogeneous \textit{self-consumption} profiles within relatively small groups are expected to mitigate grid congestion and enhance the overall efficiency of national energy systems. However, integrating these entities into existing energy systems is inherently complex, as they can significantly alter both a country's energy mix and the economic equilibrium of institutional stakeholders, including generators, distributors, and retailers.



\begin{figure}[H]
     \centering
     \begin{subfigure}[b]{0.45\textwidth}
         \centering
         \includegraphics[width=\textwidth]{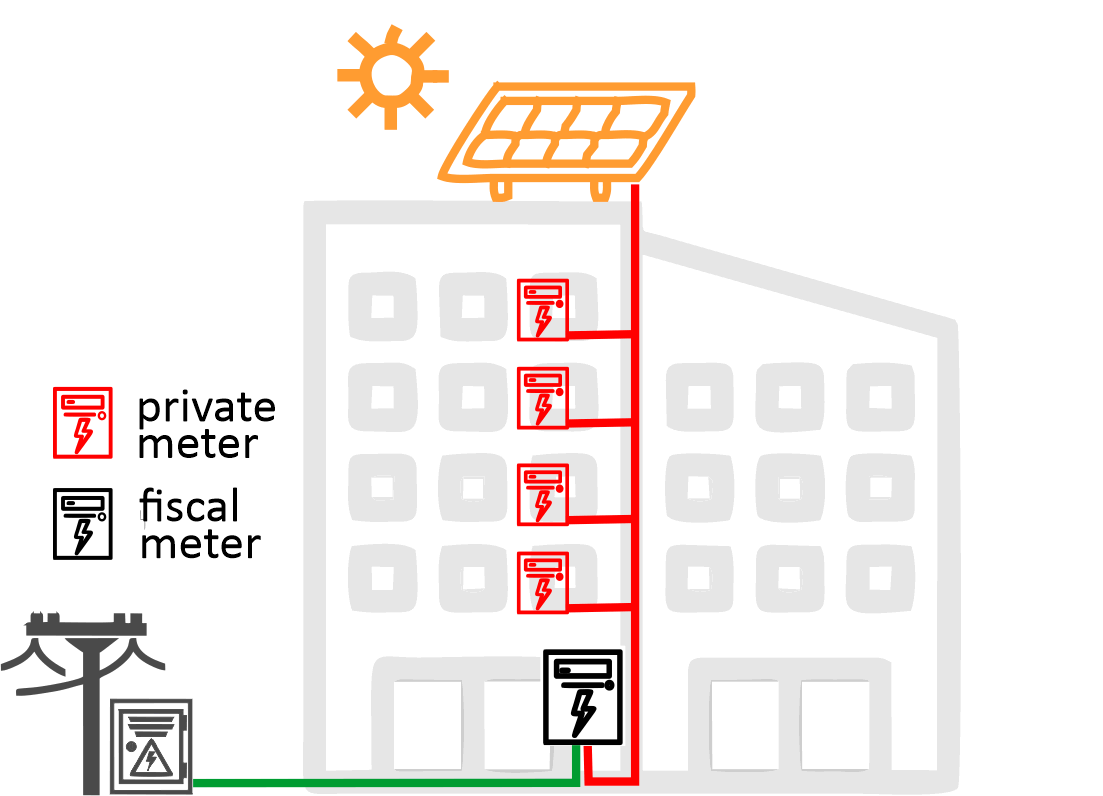}
         \caption{Physical framework}
         \label{fram1}
     \end{subfigure}
     \hfill
     \begin{subfigure}[b]{0.45\textwidth}
         \centering
         \includegraphics[width=\textwidth]{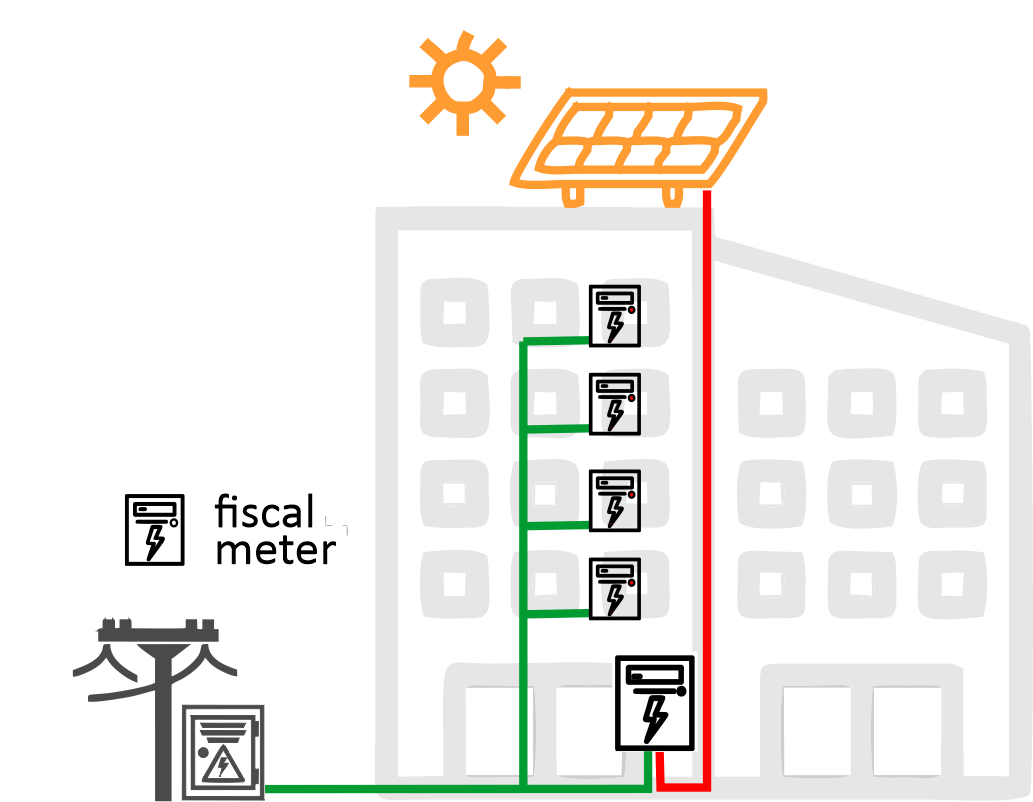}
         \caption{Virtual framework}
         \label{fram2}
     \end{subfigure}
        \caption{Difference between physical and virtual REC frameworks}
        \label{framew1}
\end{figure}

The implementation of EU directives across Member States is progressing at varying speeds and in accordance with different national regulations. A key decision faced by national regulators is whether to adopt a physical or virtual framework for Renewable Energy Communities (RECs).

For instance, Italy was among the first countries to implement the EU directives on RECs, establishing a sophisticated virtual framework for energy self-consumption \cite{IT1,IT2,IT3}. This pioneering approach has the potential to influence and inspire other EU countries seeking to leverage RECs for the energy transition.

Figure \ref{framew1} illustrates the distinction between physical and virtual frameworks, using the simplest case of an REC: a group of households within the same building equipped with a photovoltaic (PV) system. 

In the physical framework (Figure \ref{fram1}), all households are directly connected to the PV system and equipped with private meters to track individual consumption. When PV generation at time $t$ exceeds the building's aggregate energy demand, \textit{self-consumption} occurs at no cost. Only the surplus energy is recorded by a fiscal meter and sold to the grid. Conversely, if PV generation is insufficient, the additional energy demand is supplied by the grid, measured by the fiscal meter, and allocated to each household based on private meter readings.

According to the virtual framework shown in Figure \ref{fram2}, the REC members do not directly consume the energy generated with their plants (with some negligible exceptions). Under the virtual scheme, all the generation of a REC, as measured by the central fiscal meter, must be offered to the market. As for the selling price, since the REC are going to be relatively small players in size, a central authority (e.g., GSE in Italy), which acts as guarantor and promoter of the country's sustainable development, is established as a natural broker, guaranteeing 
a selling price calculated as an average of the gross price of electricity updated periodically. 
To satisfy their energy needs, the members entering a virtual REC must, therefore, continue to buy electricity from their retailers, their consumption remaining under the control of individual fiscal meters. 
This is why, for them, self-consumption occurs only in a "virtual way":
the central authority, by reading both the central and the individual meters, establishes, at every hour, how much energy has been produced by the PV system and collectively consumed by the members of the REC.
Under the virtual framework, the self-consumption of a REC is then defined on hourly basis as the minimum between the collective consumption and the renewable generation.

Since the members of a REC continue purchasing all the energy they consume while selling all the energy generated by their plants, the primary benefit of joining a REC lies in the incentive mechanism. 

More specifically, virtual RECs benefit from an incentive proportional to the total self-consumption accumulated over a given period, such as one year. While the virtual framework is undoubtedly more complex than the physical one, it offers a significant advantage in mitigating energy poverty within society. Given the definition of self-consumption, even members who join a REC solely as consumers—without contributing to the initial investment in generation plants—play a beneficial role. Their participation supports self-consumption and enhances REC revenues, particularly during peak renewable generation periods when production exceeds demand. This scenario commonly arises in RECs relying on PV generation during midday hours.

However, a critical issue, which, to the best of our knowledge, has not yet been addressed in the literature, is the development of a fair and effective mechanism for distributing the incentive among the members of a virtual REC.


As intuitively expected, the complexity of defining a fair incentive-sharing rule increases with the number of REC members, the diversity of their stakes, and their varying consumption profiles. A particularly critical factor is the stake of those members who invest in the generation infrastructure, as they must determine the optimal capacity for the REC.

The viability of a REC depends crucially on the establishment of an equitable sharing rule for the self-consumption incentive, which directly influences members' economic decisions. This rule affects both the willingness of potential members to join and the retention of existing members, as it shapes their expected benefits.

The relevance of our analysis is twofold. Firstly, at the individual level, the decision to enter or remain in a virtual REC involves a complex economic evaluation, where the incentive-sharing mechanism interacts with investment choices under uncertainty. Secondly, from an environmental policy perspective, understanding the conditions that support or hinder the long-term acceptance and stability of RECs is essential for assessing their potential impact on the energy transition.



This paper is organized as follows. Section 2 reviews the relevant bibliography 
and the studies of RECs from the optimal design and the community value point of view. Section 3 analyzes the optimal installation of different REC members in the absence of incentives. In Section 4, we present a 
Stackelberg game combined with a Nash competitive game model for the members of a REC in the presence of an incentive based on self-consumption. 
In Section 5, we apply the model to real Italian data and present the results for two particular cases. Section 6 concludes.

\section{Bibliography review} 

A comprehensive review of RECs can be found in the monograph \cite{RECbook}, where the authors examine the topic from multiple perspectives, including regulatory, legal, economic, and power system coordination challenges. The analysis extends beyond the European context to include North America, covering REC implementations in Spain, Italy, Canada, the U.S., Germany, Switzerland, France, Brazil, Denmark, Bulgaria, and Greece.

In this paper, we specifically focus on the equilibrium of RECs operating under the virtual framework, with particular attention to the Italian case, where the most advanced implementations have been developed. A review of the status of energy communities in Italy is available in \cite{RI}, while recent studies addressing management and legal aspects can be found in \cite{ECIt1, ECIt3, ECIt4}. Additionally, the impact of this new type of market participant on the power grid is analyzed in \cite{ECIt2}.

Various modeling tools have been applied in the literature to address optimal investment strategies, power grid sizing, and fair incentive-sharing mechanisms. Among these, game-theoretic approaches have been widely used, as they provide a structured framework for situations where multiple participants must cooperate to maximize collective benefits.

In the electricity sector, game-theoretic methods were initially applied to peer-to-peer energy trading in smart grids and distribution systems, demonstrating improved performance in such mechanisms \cite{P2PStak, P2Pstak2, P2PGame, WCRS}. More recently, these techniques have been employed to study the optimal investment and management of RECs. By leveraging game theory to fairly distribute community value and optimize investment in power generation and energy storage, several studies report cost reductions and enhanced benefits for REC members \cite{CGS}. The literature shows a growing trend toward using cooperative game theory for these purposes \cite{VEC, CGEC, MMM, MCA, CGS}, with occasional applications of Stackelberg games \cite{CSG2}.

In \cite{CGS}, the authors propose a coalition game to distribute value among energy community members and demonstrate that joining a REC consistently leads to lower consumption costs for prosumers.
The study in \cite{VEC} applies cooperative game theory to assess REC viability in terms of investment, considering a REC composed of $n$ households willing to invest in photovoltaic panels. The authors employ the Shapley value to fairly distribute gains among members and establish criteria for coalition stability.

The work in \cite{MMM} focuses on optimizing the investment portfolio of a REC, accounting for variations in energy consumption and generation potential among members. Initially, the problem is solved under a centralized framework where a single entity determines the optimal REC size. In a second step, cooperative game theory is applied to fairly distribute the incentive received by the community, using the Shapley value to allocate benefits and maximize the economic value of the REC. This optimization approach is extended in \cite{ECIt}, where the design and operation of generation assets are optimized to maximize community cash flows. Similarly, \cite{MCA} introduces a cooperative game-theoretic method for fair redistribution of benefits from community-owned energy assets.

However, cooperative approaches may not fully capture the dynamics among REC members, as collaboration is not always guaranteed. This limitation is discussed in \cite{BCCGEC}, where the authors address prosumer integration and decentralized decision-making using blockchain mechanisms alongside cooperative games. The success of peer-to-peer trading models within energy communities can be affected by coordination challenges. A notable study that considers the strategic behavior of members before joining a REC is \cite{GTMLEC}, which proposes an incentive mechanism to promote energy production and self-consumption among individualistic players.

\subsection{Main Contribution}

We propose a simplified model of a Renewable Energy Community consisting of a biogas producer and a representative household. This configuration is relevant both in agricultural areas, where biogas plants treat livestock manure, and in urban settings, where they process organic waste. In our framework, the biogas plant supplies power, while the household contributes both demand and additional power production. The household is modeled as a representative entity aggregating multiple similar agents, making the same optimal decisions within the REC.

This work makes several contributions. First, we develop a novel model to capture the complex decision-making process of REC members, who must simultaneously determine a fair sharing rule for the self-consumption incentive to ensure community viability and make optimal investment decisions. These two objectives are inherently interdependent, making their joint optimization a crucial aspect of REC stability.

To explicitly account for these interactions, we introduce a bi-level optimization framework. At the upper level, a REC coordinator acts as the leader in a Stackelberg game, optimizing the "opportunities balance" between the two members using a Nash bargaining criterion. At the lower level, the biogas producer and the household engage in a Nash game, where they maximize their expected profits by selecting the optimal plant capacities, given the sharing rule set by the coordinator.

The approach taken in this work significantly differs from existing literature on RECs. While most studies assume cooperative behavior among members, we adopt a competitive strategic perspective in which members act in their own self-interest when joining the REC. This assumption is more realistic, as solutions from cooperative game theory often result in asymmetric benefits, where some members gain an advantage by acting strategically while others continue to cooperate. Such situations pose a significant risk to the long-term viability of RECs.

Finally, the analytical formulation presented in this study enhances the interpretability of the results, providing a structured mathematical foundation for understanding strategic interactions within RECs.

\section{Optimal individual investment decision}

In this section, we introduce the mathematical model describing the optimal investment problem each member would solve individually. More precisely, we derive the expected profit that two potential REC members would obtain by investing in renewable generation capacity without forming a REC, i.e., remaining independent entities. These results serve as a benchmark for comparison with the profits they would achieve by joining a REC.

Both decision-makers determine their investment at time $t=0$, under the assumption of an infinite time horizon, reflecting the indefinite lifespan of the installed plants.

Individuals who simultaneously act as energy producers and consumers are commonly referred to as \textit{prosumers}. At any time $t$, they self-consume the energy generated by their plants and sell any surplus to the market, while purchasing electricity when self-generation is insufficient. Investing in a generation plant lowers energy costs for stand-alone prosumers, though they do not benefit from the self-consumption incentives available to REC members.

Renewable energy investments are subject to multiple sources of risk. To represent this uncertainty, we define a complete filtered probability space $(\Omega, \mathcal{F}, (\mathcal{F}_t)_{t \in [0,+\infty)}, \mathbb{P})$ satisfying the usual conditions, where four correlated Brownian motions, $W_c$, $W_v$, $W_p$, and $W_d$, are introduced. Among these, the correlation $\rho_c := \mathrm{Corr}(W_d, W_c)$ is particularly relevant in our analysis.

The biogas producer has a total gas production capacity $K_g$ (in m$^3$), with an hourly output of $b K_g$, where $b>0$ is the MW/m$^3$ conversion factor. The producer sells this gas at the spot market price $(P^p(t))_{t \geq 0}$ (in \euro $/$MWh), which follows a geometric Brownian motion with initial value $p$:

\begin{equation}
P^p(s) = p e^{\mu_p s + \sigma_p W_p(s)},
\end{equation}

\noindent where $\mu_p \in \mathbb{R}$ and $\sigma_p > 0$. 

The biogas producer can install a Gas-to-Power (G2P) turbine with capacity $y_b \leq \theta_b$, where $\theta_b$ represents the maximum feasible capacity. This turbine converts gas into electricity, which is sold at the spot electricity price $(X_v^{x_v}(t))_{t \geq 0}$. We model this price as a geometric Brownian motion with initial value $x_v$:

\begin{equation}
X^{x_v}_v(s) = x_v e^{\mu_v s + \sigma_v W_v(s)},
\end{equation}

\noindent where $\mu_v \in \mathbb{R}$ and $\sigma_v > 0$. 

Installing the turbine reduces the amount of gas available for direct sale. Specifically, the gas output decreases to $(b K_{g} - y_b)$, while electricity production increases by $\gamma y_b$, where $\gamma$ denotes the G2P efficiency factor\footnote{Typically, $\gamma \simeq$ 0.5--0.6 for gas-fired power plants}. 

The biogas producer's profit functional is thus given by:

\begin{equation} \label{J0b}
J^0_b(x_v,x_c,p,d,y_b,y_h) :=  \mathbb{E}\left[ \int_{0}^{\infty} e^{-r s}  X^{x_v}_v(s)  \gamma y_b  ds  \right]  
+  \mathbb{E}\left[ \int_{0}^{\infty} e^{-r s}  P^p(s)(b K_{g} -  y_b) ds \right]   - c_b y_b\text{,}
\end{equation}

\noindent where $c_b$ represents the cost coefficient (\euro/MW) of installing a turbine with capacity $y_b$, and $r>0$ is the discount rate reflecting the cost of capital. Given the intrinsic risk exposure of both players, $r$ is assumed to be significantly higher than a risk-neutral rate.

To meet its electricity demand, the household purchases energy from the market at the spot price $(X_c^{x_c}(t))_{t \geq 0}$, which follows a geometric Brownian motion:

\begin{equation}
X^{x_c}_c(s) = x_c e^{\mu_c s + \sigma_c W_c(s)},
\end{equation}

\noindent where $\mu_c \in \mathbb{R}$ and $\sigma_c > 0$. The household’s instantaneous power demand $(D^d(t))_{t \geq 0}$ is also modeled as a geometric Brownian motion:

\begin{equation}
D^d(s) = de^{\mu_d s + \sigma_d W_d(s)},
\end{equation}

\noindent with $\mu_d \in \mathbb{R}$ and $\sigma_d > 0$. For consistency, we express demand $D^d$ in MW.

The choice of geometric Brownian motion to model electricity and gas prices, as well as household demand, is motivated by analytical tractability. While mean-reverting processes more accurately reflect empirical energy price and demand patterns—preventing variance from diverging to infinity in the long run—we mitigate this limitation by discounting future values and imposing Assumption \ref{Ass1}. As a result, variance growth is not an issue, and our findings remain valid. Moreover, unlike time-dynamic optimization problems, where mean-reverting processes significantly influence trading and operational decisions—particularly for the biogas producer—our framework focuses on decision-making at time zero. Since our analysis is centered on initial conditions, the distinction between mean-reverting and geometric Brownian motion processes is expected to have a limited impact on our main conclusions.

The household can invest in photovoltaic panels with capacity $y_h \leq \theta_h$ to reduce energy costs by selling surplus production at the electricity spot price $X_v$, where $\theta_h$ is typically constrained by the household's budget.
The household’s profit functional is given by:

\begin{equation} \label{J0h}
J^0_h(x_v,x_c, p,d,y_b,y_h) :=  \mathbb{E}\left[ \int_{0}^{\infty} e^{-r s}  X_v^{x_v}(s) \alpha y_h  ds \right]  
- \mathbb{E}\left[ \int_{0}^{\infty} e^{-r s} X_c^{x_c}(s) D^d(s) ds \right]  - c_h y_h,
\end{equation}
\noindent where $\alpha$ represents the PV panel's equivalent power production factor\footnote{Typically, $\alpha$ is related to the "equivalent hours per year": if a PV plant operates at peak-level equivalent for $M$ hours annually, then $\alpha = \frac{M}{8760}$, with $M \in [1000,1500]$ depending on location.} and $c_h$ is the installation cost per MW. 
\color{b}
While constant photovoltaic generation is an evident simplification, generalizing this assumption is not expected to modify the results of this analysis. Indeed, local weather conditions and solar lighting can be assumed to be independent of the national price of electricity. 
Given that we consider the expectation of that integral, $\alpha y_h$ can be safely seen as the constant long run average rate of the photovoltaic generation process.

Note that both plant installations are assumed to operate at full capacity at all times, with $y_b$ representing the output of the G2P turbine and $\alpha y_h$ the output of the PV panels. Thus, in this framework, installed capacities coincide with generation rates.


We now determine the optimal installation levels for both the biogas producer and the household in the absence of REC incentives. To compute $J^0_b$ and $J^0_h$, we impose the following assumption.

\begin{assumption}\label{Ass1}
The parameters
\begin{equation*}
r_v := r - \mu_v - \frac{\sigma^2_v}{2}, \quad r_p := r - \mu_p - \frac{\sigma^2_p}{2}, \quad r_c := r - \mu_c - \frac{\sigma^2_c}{2}, \quad r_d := r - \mu_d - \frac{\sigma^2_d}{2}, \quad r_{cd} = r_c + r_d - r  -  \rho_c \sigma_c \sigma_d.
\end{equation*}
are strictly positive.
\end{assumption}

These quantities represent discount rates adjusted for the expected growth of the corresponding stochastic processes. Under Assumption \ref{Ass1}, it follows that:

\begin{equation*}
\mathbb{E}[e^{-r s} X^{x_v}_v(s)] = x_v e^{-r_v s}, \quad 
\mathbb{E}[e^{-r s} P^p(s)] = p e^{-r_p s}, \quad 
\mathbb{E}[e^{-r s} X_c^{x_c}(s) D^d(s)] = x_c d e^{- r_{cd} s},
\end{equation*}

\noindent allowing us to rewrite \eqref{J0b} and \eqref{J0h} as:

\begin{eqnarray*}
J^0_b(x_v,x_c,p,d,y_b,y_h) &=& 
 \gamma \frac{x_v y_b}{r_v} + \frac{p (b K_{g} -  y_b)}{r_p} - c_b y_b = y_b g_b + \frac{p b K_{g}}{r_p}, \\
J^0_h(x_v,x_c, p,d,y_b,y_h) &=& \alpha \frac{x_v y_h}{r_v} - \frac{x_c d}{r_{cd}} - c_h y_h = y_h g_h - \frac{x_c d}{r_{cd}}.
\end{eqnarray*}

The quantities
\begin{equation} \label{gbh}
g_b := \gamma \frac{x_v}{r_v} - \frac{p}{r_p} - c_b, \qquad
g_h := \alpha \frac{x_v}{r_v} - c_h
\end{equation}
represent the discounted net unit profits per MW for the biogas producer (from G2P investment) and the household (from PV generation) over an infinite time horizon.

Notably, $J^0_b$ and $J^0_h$ correctly reflect the independence of the two agents, as $J^0_b$ does not depend on $y_h$, and $J^0_h$ does not depend on $y_b$. Since both profit functions are linear in $y_b$ and $y_h$, the following result follows directly.

\begin{lemma} \label{J0}
In the absence of incentives, the optimal installation levels for the biogas producer and household are given by:

\begin{equation*}
y_b^* = \begin{cases}    
    \theta_b & \mbox{ if } g_b \geq 0,\\
    0 & \mbox{ if } g_b < 0,\\
\end{cases} 
\quad \text{and} \quad
y_h^* = \begin{cases}    
    \theta_h & \mbox{ if } g_h \geq 0,\\
    0 & \mbox{ if } g_h < 0.\\
\end{cases}
\end{equation*}
\end{lemma}

\begin{remark} \label{noincentive}
This result follows the classical "\textit{all-or-nothing}" solution of linear investment optimization. Specifically, if the net unit profit $g_b$ (for the biogas producer) or $g_h$ (for the household) is non-negative, the agent installs the maximum possible capacity; otherwise, no installation occurs.

For the household, the expected revenue is given by $\frac{x_v}{r_v}$, the current electricity selling price $x_v$ divided by the net discount rate $r_v = r - \mu_v - \sigma^2_v/2$. If the marginal installation cost $c_h$ exceeds this expected revenue, the optimal installation is zero. Similarly, for the biogas producer, expected revenue is reduced by the gas price term $\frac{p}{r_p}$. If $c_b$ exceeds expected revenues, no investment in G2P capacity occurs.
\end{remark}

\color{b}

\section{A model for Virtual Renewable Energy Communities} 

\subsection{Optimal sharing rule and capacity decisions within a virtual REC}

According to the Virtual Framework discussed in the Introduction, a REC receives an incentive proportional to its self-consumption, which is then distributed among its members. In our case, both the biogas producer and the household may revise their installation decisions from the previous section in light of these incentives. As we will show, investment decisions and the incentive-sharing rule interact in a nontrivial manner.

More specifically, if the quantities $g_b$ and $g_h$ are nonnegative for the two players as stand-alone prosumers, the additional incentives from joining a REC would simply reinforce their decision to invest up to the maximum capacity. However, if either $g_b$ or $g_h$ is negative, the incentives may shift a zero-investment decision into a positive investment. In this case, as we will demonstrate, the optimal investment is not necessarily equal to the maximum capacity limit.

To quantitatively assess this effect, we now describe the self-consumption incentive in detail. Both members contribute to the total power generated by the REC: the household installs photovoltaic panels providing $y_h$ units of power, while the biogas producer installs P2G turbines generating $y_b$ units of electricity from gas conversion. For virtual RECs, self-consumption at any time $t$ is typically defined as:

\begin{eqnarray} \label{qt}
q(t) = \min(D^d(t), y_h + y_b).
\label{selfc}
\end{eqnarray}

Self-consumption plays a central role in the incentive system adopted for virtual RECs. Increasing self-consumption depends not only on the generation capacity $y_h + y_b$ but also on the aggregate demand of the REC, $D^d(t)$. This structure supports the mitigation of \textit{energy poverty}, as individuals who join the REC without making an initial investment still contribute by increasing aggregate demand, thereby enhancing the REC’s overall self-consumption.

We assume that a central authority rewards a REC with a flat tariff $Z$ for every unit of self-consumed energy over a period $[0,\tau]$, where $\tau$ is a random variable representing the duration of the incentive scheme, accounting for the possibility that the policy may be discontinued at an uncertain future time.

Since the REC operates as a single legal entity and receives the self-consumption incentive collectively, we assume that a designated coordinator is responsible for determining its distribution among members.
\color{b}

We model this situation as a Stackelberg game, where the coordinator acts as the strategic leader and the two REC members (the biogas producer and the household) act as followers. More precisely, the coordinator optimizes a Nash bargaining objective, anticipating the optimal responses of the followers. If $\beta \in (0,1)$ represents the share of the incentive allocated to the household, with $1-\beta$ assigned to the biogas producer, the coordinator solves the following optimization problem:

\begin{eqnarray}
\max_{\beta \in (0,1)}  \left( J_h(x_v,x_c,p,d,y_b,y_h, \beta) - d_h \right)\left( J_b(x_v,x_c,p,d,y_b,y_h,\beta) - d_b \right),
\label{NBarg}
\end{eqnarray}

where $J_h(x_v,x_c,p,d,y_b,y_h,\beta)$ and $J_b(x_v,x_c,p,d,y_b,y_h,\beta)$ denote the profits of the household and biogas producer upon joining the REC, while $d_h$ and $d_b$ are the \textit{disagreement points}, corresponding to their respective profits if they remain independent. Since the objective function in \eqref{NBarg} is the product of two terms, it reaches zero whenever either term is zero. This structure naturally prevents highly asymmetric solutions where one member benefits disproportionately at the expense of the other.

As required in a Stackelberg game, the coordinator acts strategically. To solve problem (\ref{NBarg}), it must compute $J_h$ and $J_b$ as best responses of the followers $(y_h^*,y_b^*)$, given its choice of $\beta$.

The disagreement points $d_h$ and $d_b$ represent the profits of the agents when acting independently. Using Lemma \ref{J0}, they are given by:

\begin{eqnarray}
d_h := J^0_h(x_v,x_c,p,d,y_h^*,y_b^*), \qquad  d_b := J^0_b(x_v,x_c,p,d,y_h^*,y_b^*).
     \label{dishb}
\end{eqnarray}
We now define the profit functions $J_h$ and $J_b$ for the household and the biogas producer. The biogas producer's profit is expressed as:
\begin{eqnarray*}
J_b(x_v,x_c,p,d,y_b,y_h, \beta) := J^0_b(x_v,x_c,p,d,y_b,y_h) + (1- \beta) Z w(y_h,y_b,d),
\end{eqnarray*}
where $J^0_b$ represents the biogas producer's profit in the absence of incentives, as defined in Equation \eqref{J0b}. The term $(1- \beta) Z w(y_h,y_b,d)$ accounts for the incentive received by the biogas producer, where $w(y_h,y_b,d)$ is the expected self-consumption cumulated by the REC over its lifetime:

\begin{eqnarray} \label{w(d)}
w(y_h,y_b,d) := \mathbb{E}\left[\int_{0}^{\tau} e^{-r s} q(s)ds \right].
\label{aux1}
\end{eqnarray}

Similarly, the household's profit is given by:

\begin{eqnarray*}
J_h(x_v,x_c, p,d,y_b,y_h, \beta) := J^0_h(x_v,x_c, p,d,y_b,y_h) + \beta Z w(y_h,y_b,d),
\end{eqnarray*}

where $J^0_h$ is the household's profit without incentives, defined in Equation \eqref{J0h}. 

For each fixed $\beta$, the members of the REC seek Nash equilibrium points $(y_h^*,y_b^*)$ satisfying:

\begin{gather} \label{NasDef}
    \begin{cases}
    J_h(x_v,x_c,p,d,y_h,y_b^*,\beta) \leq J_h(x_v,x_c,p,d,y_h^*,y_b^*, \beta) & \forall y_h \in [0, \theta_h],\\
    J_b(x_v,x_c,p,d,y_h^*,y_b, \beta) \leq J_b(x_v,x_c,p,d,y_h^*,y_b^*, \beta) & \forall y_b \in [0, \theta_b].
    \end{cases}
\end{gather}

In principle, the two members could cooperate by independently agreeing on a "reasonable" value for $\beta$. However, such coordination is uncommon in practice. Instead, our approach assumes a designated coordinator determines a "fair" sharing rule, allowing the two members to compete in optimizing their respective installation capacities within a structured framework.

\color{b}

\begin{figure}[H]
    \centering
    \includegraphics[scale = 1.2]{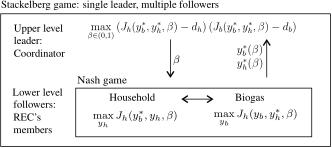}
    \caption{Representation of the bilevel problem, with a Nash game nested within a Stackelberg game}
    \label{BilevelProb}
\end{figure}

As is common in bilevel optimization problems, we solve our leader-follower framework using a two-step procedure. First, we determine the Nash equilibrium of the competitive installation game between the REC members for a given $\beta$. Then, using the resulting optimal capacities $y_h^*(\beta)$ and $y_b^*(\beta)$, we solve the leader's bargaining problem, as illustrated in Figure \ref{BilevelProb}.

\subsection{Lower-level problem: Nash equilibrium between the REC members}
\subsubsection{Preliminary results}

In this section we compute the expected value of the household and biogas producer profits, by noticing that the cumulative incentive function $w$ in Equation \eqref{w(d)}, 
under suitable assumptions, 
can be computed using the Feynman-Kac formula \cite[Remark 3.5.6]{pham}. 

Let us begin with the following lemma.
\begin{lemma} Assume that $r_d >0$. Assume furtherly that $\tau$ is independent of $W_d$ and exponentially distributed with parameter $\lambda > 0$. Then the function $w$, defined in Equation \eqref{w(d)}, is of class $C^2(({\mathbb R}^+)^3)$, and is equal to
\begin{gather} \label{w(y,d)}
w({y_h,y_b,}d) :=  \begin{cases} 
B_1 (y_h + y_b)^{1-m_2}d^{m_2} + \frac{d}{r_d {+ \lambda}} & d < y_h + y_b\\
C_1(y_h + y_b)^{1-m_1}d^{m_1} + \frac{y_h +y_b}{r {+ \lambda}} & d \geq y_h + y_b,
\end{cases}
\end{gather}
with $B_1$ and $C_1$ negative constants given by
\begin{equation} \label{B1C1}
B_1 = \frac{m_1 \mu_d + m_1 \sigma^2_d/2 - r {- \lambda}}{(r {+ \lambda})(r_d {+ \lambda})(m_2 - m_1)}, \qquad
C_1 = \frac{m_2 \mu_d + m_2 \sigma_d^2 /2 - r { - \lambda}}{(r { + \lambda}) (r_d { + \lambda})(m_2 - m_1)} 
\end{equation}
and $m_1 < 0 < m_2$ given by
$$ m_1 := \frac{-\mu_d - \sqrt{\mu_d^2 + 2 (r { + \lambda}) \sigma_d^2}}{\sigma_d^2}, \qquad m_2 := \frac{-\mu_d + \sqrt{\mu_d^2 + 2 (r { + \lambda}) \sigma_d^2}}{\sigma_d^2} $$
\label{lemmamin}
\end{lemma}

\begin{proof}
See Appendix.
\end{proof}

\begin{remark}
In the previous lemma, the inequality $m_1 < 0 < m_2$ holds independently of the sign of $\mu_d$. Moreover, it is not difficult to prove that these general inequalities can be strengthened, as we have that $m_1 < - 1$ if $\mu_d > 0$ and $m_2 > 1$ if $\mu_d < 0$. Observe also that the function $J_h$ is concave in $y_h$ and $J_b$ is concave in $y_b$. Moreover, one can verify by direct computation that $J_h$ is $C^1$ with respect to $y_h$ and $J_b$ is $C^1$  with respect to $y_b$.
\end{remark}


We now examine the competitive game of power installation between the community members, seeking Nash equilibria. Since each player's optimal decision depends on the best response of the other, we derive explicit analytical expressions for their strategies, ensuring an equilibrium characterization.  
\color{b}

\subsubsection{Household's best response}

Let us begin with the household's best response for a fixed strategy $y_b \in [0, \theta_b]$ of the biogas producer. We want to find $y_h^*$ such that

\begin{eqnarray}
J_h(x_v,x_c,p,d,y_h^*, y_b, \beta) = \max_{y_h \in [0, \theta_h]} J_h(x_v,x_c,p,d,y_h, y_b, \beta).
\label{maxjh}
\end{eqnarray}

\begin{proposition}
For $\beta \in (0,1)$ and $y_b \in  [0, \theta_b]$ fixed, the best response $y_h^*$ for the household to the strategy $y_b$ of the biogas producer is

\begin{gather}
    y_h^* = \begin{cases} \theta_h, & g_h 
    \geq 0, \\
    0, & g_h + \frac{Z \beta }{r+ \lambda}  \leq 0,\\
     \max \left( \min \left( \displaystyle d\left( \frac{\beta}{\beta_h}\right)^{1/m_2} - y_b , \theta_h \right) , 0 \right)  & g_h < 0 < g_h + \frac{Z \beta }{r+ \lambda}  \wedge \beta > \beta_h, \\
     \max \left( \min \left(  d\left( \displaystyle \frac{  \frac{g_h}{\beta_h}   + \frac{Z}{r+ \lambda}  }{\frac{g_h}{\beta}  + \frac{Z}{r+ \lambda} }\right)^{1/m_1} - y_b , \theta_h \right) , 0 \right) & g_h < 0 < g_h + \frac{Z \beta }{r+ \lambda}  \wedge \beta \leq \beta_h,
    \end{cases}
    \label{hbestr}
\end{gather}

\noindent where $\beta_h$ is defined as
\begin{equation} \label{betah}
\beta_h : = \displaystyle \frac{- g_h}{Z (C_1 (1 - m_1) + \frac{1}{r + \lambda})} = \frac{- g_h}{Z B_1 (1 - m_2)}. 
\end{equation}
and it is strictly positive if and only if $g_h < 0$.

In addition, if $\beta_h < \beta$, then the aggregate installation exceeds the power demand, i.e., $y_h^* + y_b > d$. On the other hand, if $\beta_h \geq \beta$, then the aggregate installation does not exceed the community demand, i.e., $y_h^* + y_b \leq d$. The equality is verified when $\beta = \beta_h$. 
\label{proph}
\end{proposition}

{\begin{remark} \label{besthousehold}
As in the no-incentive case, discussed in Remark \ref{noincentive}, the interpretation of this result is straightforward. This time, the reward functional is strictly concave, due to the presence of the incentive, which adds a concave nonlinearity to the linear functional without incentives. This result in the existence of a unique maximizer $y_h^*$: since this maximizer must to be found in the compact set $[0,\theta_b]$, several cases arise. If the net unitary gain $g_h$ is non-negative, then the optimum is reached by installing the maximum possible power: in fact, this was already optimal without the incentive and remanis so in this case. Conversely, a zero optimal installation now is optimal only if $g_h$ (which is the net gain {\em without incentives}), summed with the expected value over time of the household's incentive share $\beta \frac{Z}{r+\lambda}$, is still non-positive: this would mean that the household's incentive share is not sufficient to make the total gain of the household positive. Finally, we have possibly internal solutions $y_h^* \in (0,\theta_h]$ when the net gain $g_h$ is negative without incentive but becomes positive when summed with the household's incentive share $\beta \frac{Z}{r+\lambda}$. Since the functional form of the incentive in Equation \eqref{w(y,d)} assumes two analytic expressions according to the values of $d$ and $y_h + y_b$, here we have two functional forms for the inner maximum point $y_h^*$: it turns out that the choice between these two forms is uniquely determined by whether $\beta$ is greater than or less than a given constant $\beta_h$.
\end{remark}
}


 \subsubsection{Biogas producer's best response}
 
Let us move to the biogas producer's best response problem. For a fixed $y_h \in [0, \theta_h]$, we want to find $y_b^*$ such that

\begin{eqnarray}
J_b(x_v,x_c,p,d,y_h, y_b^*, \beta) = \max_{y_b \in [0, \theta_b]} J_b(x_v,x_c,p,d,y_h, y_b, \beta).
\label{maxjb}
\end{eqnarray}

\begin{proposition}
For $\beta \in (0,1)$ and $y_h \in  [0, \theta_h]$ fixed, the best response $y_b^*$ for the biogas producer to the strategy of the household is

\begin{gather}
    y_b^* = \begin{cases} \theta_b, & g_b \geq 0\\
    0, & g_b + \frac{Z(1 - \beta)}{r+ \lambda}  \leq 0\\
    \max \left( \min \left(  d \left( \frac{1 - \beta}{1 - \beta_b } \right)^{1/m_2} - y_h , \theta_b \right) , 0 \right), & g_b < 0 < g_b + \frac{Z(1- \beta) }{r+ \lambda}  \wedge \beta < \beta_b\\
     \max \left( \min \left( \displaystyle d \left( \frac{\frac{g_b}{1 - \beta_b} + \frac{Z}{r+ \lambda}}{\frac{g_b}{1 - \beta} + \frac{Z}{r+ \lambda}}  \right)^{1/m_1} - y_h , \theta_b \right) , 0 \right), & g_b < 0 < g_b + \frac{Z(1- \beta) }{r+ \lambda}  \wedge \beta \geq \beta_b
    \end{cases}
    \label{bbestr}
\end{gather}
where $\beta_b$ is defined as
\begin{equation} \label{betab}
\beta_b : = 1 + \frac{- g_b}{ Z B_1 (m_2 - 1)} = \frac{g_b}{Z ( C_1 (1 - m_1) + \frac{1}{r+ \lambda})} +1 .
\end{equation}
We have that $1 - \beta_b > 0$, if and only if, $g_b < 0$. 
In addition, if $\beta_b > \beta$, then the aggregate installation exceeds the power demand, i.e., $y_h + y_b^* > d$. On the other hand, if $\beta_b \leq \beta$, then the aggregate installation does not exceed the community demand, i.e., $y_h + y_b^* \leq d$. The equality holds when $\beta = \beta_b$. 
\label{propb}
\end{proposition}

\begin{remark} \label{bestbiogas}
As in the case of the household, discussed in Remark \ref{besthousehold}, the interpretation of this result is also straightforward. The reward functional is again strictly concave due to the presence of the incentive, thus again the maximizer  $y_b^* \in [0,\theta_b]$ is unique, and several cases arise. 
If the net unitary gain $g_b$ is non-negative, then the optimum is reached by installing the maximum possible power $\theta_b$, which was also optimal without the incentive. 
In analogy with Remark \ref{besthousehold}, a zero optimal installation now is optimal, only if the net gain without incentive $g_b$ summed with the expected value over time of the biogas producer's incentive share, which is now $(1 - \beta) \frac{Z}{r + \lambda}$, is still non-positive, as in this case the biogas producer's incentive share is not sufficient to make the total gain positive. 
Finally, we have possibly internal solutions $y_b^* \in (0,\theta_b]$ when the net gain $g_b$ is negative without incentive but becomes positive when summed to the biogas producer's incentive share $(1 - \beta) \frac{Z}{r + \lambda}$. 
Again, we have two functional forms for the inner maximum point $y_b^*$, with the choice between these two forms being uniquely determined by $\beta$ being greater or less than a given constant $\beta_b$, which is such that $1 - \beta_b$ has the opposite sign of $g_b$. 
\end{remark}

As previously discussed, the solution $y_h^*=0$ is consistent with the creation of a REC, since it represents the case of a household contributing to the REC's self-consumption only through its energy demand. However, for the biogas producer we let  $y_b^* \in (0,\theta_b]$, that is, we exclude the case $y_b^*=0$. Though $y_b^*=0$ is consistent with a Nash equilibrium, it is not consistent with the creation of a REC, since we assumed that the biogas producer has zero power demand. For this reason, unless the biogas producer provides the REC positive generation capacity, it would have no reason to be a member of the REC.

\subsubsection{Nash equilibria} 
\label{sectionNash}
Now, we study the case where both players simultaneously apply the best response to the other player's strategy. As we saw in the previous sections, we will have different equilibria the different parameters values. For the sake of simplicity we will suppose that $d < \theta_h + \theta_b$. 

\begin{proposition} \label{ne} For $\beta \in (0,1)$ fixed, the energy community is created if and only if $g_b + \frac{Z}{r+ \lambda} (1 - \beta) \geq 0$, $y_h^* \geq 0$ and $y_b^* > 0$, with $y_h^*$ and $y_b^*$ defined respectively by Propositions \ref{proph} and \ref{propb}. In the case where $y_h^* \in [0, \theta_h]$ and $y_b^* \in (0, \theta_b]$, the following strategies $(y_h^*, y_b^*)$ are Nash equilibria.
\begin{itemize}
    \item[1. ] If $g_h \geq 0$ , $g_b \geq 0$, then \begin{eqnarray}\label{NYC11}   y_h^* = \theta_h & \mbox{, } & y_b^* = \theta_b.
    \end{eqnarray} 
   
    \item[2. ] If $g_h + \frac{Z}{r+ \lambda} \beta \leq 0$ , $g_b \geq 0$, then
    \begin{eqnarray}\label{NYC12} y_h^* = 0 & \mbox{, } & y_b^* = \theta_b.  
    \end{eqnarray}
   
    \item[3. ] If $g_h < 0 < g_h + \frac{Z}{r+ \lambda} \beta$ , $g_b \geq 0$ and
    \begin{itemize}
        \item[3.1. ] $\beta > \beta_h$, then \begin{eqnarray}\label{NYC13} y_h^* =  \max \left\{ \min \left\{ \displaystyle d\left( \frac{\beta}{\beta_h}\right)^{1/m_2} - \theta_b , \theta_h \right\} , 0 \right\}  & \mbox{, } & y_b^* = \theta_b.
    \end{eqnarray} 
    \item[3.2. ] $\beta \leq \beta_h$, then
    \begin{eqnarray}\label{NYC14} y_h^* =  \max \left\{ \min \left\{  d\left( \displaystyle \frac{  \frac{g_h}{\beta_h}   + \frac{Z}{r+ \lambda}  }{\frac{g_h}{\beta}  + \frac{Z}{r+ \lambda} }\right)^{1/m_1} - \theta_b , \theta_h \right\} , 0 \right\}  & \mbox{, } & y_b^* = \theta_b.
    \end{eqnarray}
    \end{itemize} 
  
    \item[4.] If  $g_h \geq 0$ , $g_b < 0 < g_b + \frac{Z(1-\beta)}{r+ \lambda}$ and 
    \begin{itemize}
        \item[4.1. ]  $\beta < \beta_b$, then
    \begin{eqnarray}\label{NYC15} y_h^* = \theta_h & \mbox{, } & y_b^* = \max \left\{ \min \left\{  d \left( \frac{1 - \beta}{1 - \beta_b } \right)^{1/m_2} - \theta_h , \theta_b \right\} , 0 \right\}.
    \end{eqnarray}
    \item[4.2. ] $\beta \geq \beta_b$, then
    \begin{eqnarray}\label{NYC16} y_h^* = \theta_h & \mbox{, } & y_b^* =  \max \left\{ \min \left\{ \displaystyle d \left( \frac{\frac{g_b}{1 - \beta_b} + \frac{Z}{r+ \lambda}}{\frac{g_b}{1 - \beta} + \frac{Z}{r+ \lambda}}  \right)^{1/m_1} - \theta_h , \theta_b \right\} , 0 \right\}.
    \end{eqnarray}
    \end{itemize}
   
    \item[5.] If $ g_h + \frac{Z \beta}{r+ \lambda} \leq 0$  , $g_b < 0 < g_b + \frac{Z (1 - \beta)}{r+ \lambda}$ and
    \begin{itemize}
        \item[5.1. ] $\beta < \beta_b$, then
      \begin{eqnarray}\label{NYC17}  y_h^* = 0 & \mbox{, } & y_b^* = \max \left\{ \min \left\{  d \left( \frac{1 - \beta}{1 - \beta_b } \right)^{1/m_2}, \theta_b \right\} , 0 \right\}.
    \end{eqnarray}
    \item[5.2. ]  $\beta \geq \beta_b$, then
     \begin{eqnarray}\label{NYC18}  y_h^* = 0 & \mbox{, } & y_b^* =  \max \left\{ \min \left\{ \displaystyle d \left( \frac{\frac{g_b}{1 - \beta_b} + \frac{Z}{r+ \lambda}}{\frac{g_b}{1 - \beta} + \frac{Z}{r+ \lambda}}  \right)^{1/m_1} , \theta_b \right\} , 0 \right\}.
    \end{eqnarray}
    \end{itemize}
 \end{itemize}   
\begin{itemize}
\item[6. ] In the case $g_h \in \left(- \frac{Z \beta}{r+ \lambda},0\right)$, $g_b \in \left(- \frac{Z (1 - \beta)}{r+ \lambda},0\right)$, 
there exists a unique value $\beta=\beta_n$, defined as    
    \begin{equation} \label{betan}
    \beta_n := \frac{g_h}{g_h + g_b} = \frac{\beta_h}{1 - \beta_b + \beta_h}, 
    \end{equation}
 such that the two members find Nash equilibria $(y_h^*,y_b^*)$. 

 $\beta_n$ is always between $\beta_b$ and $\beta_h$, i.e. either $\beta_n \in (\beta_h, \beta_b)$ or $\beta_n \in (\beta_b, \beta_h)$ and, depending on which of these cases is true, we have:
  \begin{itemize}
  \item[6.1.]
    If $\beta_n \in (\beta_h, \beta_b)$, any pair $(y_h^* , y_b^*)$ such that $y_h^* \in [0, \theta_h]$ and $y_b^* \in (0, \theta_b]$, which satisfies
    \begin{eqnarray} \label{NYC2}
    y_h^* + y_b^* = d \left(  \frac{Z B_1 (m_2 - 1)   }{  g_h + g_b } \right)^{1/m_2} {= d \left(  \frac{1 - \beta_n}{1 - \beta_b} \right)^{1/m_2} }>d
    \end{eqnarray}
    \noindent is a Nash equilibrium.
  \item[6.2.]
    If $\beta_n \in (\beta_b, \beta_h)$, any pair $(y_h^* , y_b^*)$ such that $y_h^* \in [0, \theta_h]$ and $y_b^* \in (0, \theta_b]$ which satisfies
    \begin{eqnarray} \label{NYC3}
    y_h^* + y_b^* = d \left(  \frac{ Z C_1 (m_1 - 1)  }{ g_h + g_b + \frac{Z}{r+ \lambda} } \right)^{1/m_1} { = d \left( \frac{\frac{g_b}{1 - \beta_b} + \frac{Z}{r+ \lambda}}{\frac{g_b}{1 - \beta_n} + \frac{Z}{r+ \lambda}}\right)^{1/m_1} }<d
    \end{eqnarray}
    \noindent is a Nash equilibrium. 
\end{itemize}


\end{itemize}
\noindent On the other hand, if $g_b + \frac{Z}{r+ \lambda} (1 - \beta) \leq 0$, the energy community is not created, and the following decisions are Nash equilibria.
\begin{itemize}
    \item[7. ] \label{case7} If $g_h \geq 0$ , $g_b + \frac{Z}{r+ \lambda} (1 - \beta) \leq 0$, then
    \begin{eqnarray} y_h^* = \theta_h & \mbox{, } & y_b^* =0.
        \label{NE7}
    \end{eqnarray}
    \item[8. ] If $g_h + \frac{Z}{r+ \lambda} (1 - \beta) \leq 0$,  $g_b + \frac{Z}{r+ \lambda} (1 - \beta) \leq 0$, then
    \begin{eqnarray} y_h^* = 0 & \mbox{, } & y_b^* = 0.
        \label{NE8}
    \end{eqnarray}
    \item[9. ] If $g_h < 0 < g_h + \frac{Z}{r+ \lambda} (1 - \beta)$ , $g_b + \frac{Z}{r+ \lambda} (1 - \beta) \leq 0$. 
    
    As the biogas producer does not invest, the community cannot be created, and no incentive is earned by the community. Consequently, the household misses the necessary profitability and its installation is also zero:

      \begin{eqnarray} y_h^* = 0 & \mbox{, } & y_b^* = 0.
        \label{NE9}
    \end{eqnarray}
\end{itemize}
\end{proposition}

\begin{proof}
See Appendix.
\end{proof}

%
%
{
%
%

This proposition is clearly relevant both for environmental policy makers and for the players involved.
Among the above nine cases, equilibrium $6.$ (Eqs. \eqref{NYC2} and \eqref{NYC3}) is particularly relevant to policy makers, since it is where the self-consumption incentive is decisive in convincing both players to install a positive renewable capacity and, at the same time, their joint investment closely matches the expected consumption $d$ within the community. 
This is the ideal case, possibly the one in the minds of the virtual framework designers, where the incentive policy is correctly balanced and avoids different types of failures, which can occur if: i) the incentive policy is insufficient to trigger the creation of the RECs, ii) it is excessive, favoring capacity investment far beyond self-consumption, or iii) it is even unnecessary, when the decision makers would invest even in the absence of any incentive. 

Equilibrium $5.$ (Eqs. \eqref{NYC17} and \eqref{NYC18}) shows the creation of a REC. In this case, the household does not invest in any PV capacity, yet it joins the community as a pure consumer, while the biogas producer installs a positive capacity $y_b$: this installation is possible thanks to the incentive tariff $\beta$. Recall that by acting as a pure consumer, the household produces benefits also to the biogas producer: more precisely, the household consumes the biogas generation, and contributes to the self-consumption needed to receive the incentive tariff.

Equilibria $1.$, $2.$ and $3.$ analyze the case when, for one technology (biogas) the incentive is unnecessary. In this case, the biogas producer installs up to its maximum ($\theta_b$) even in the absence of any incentive. In the case of equilibrium $1.$, this is true also for PV technology.

Equilibrium $4.$ is a case where the incentive is unnecessary for PV technology.

Equilibria $7.$, $8.$ and $9.$ analyze the case when the incentive is insufficient for the biogas producer. In these cases, the energy community is not created. { In Equilibrium 7., the investment in solar panels is profitable even without incentives. In Equilibria 8. and 9., no one invests in renewable energy. 


Notice, finally, that the PV technology (in equilibrium $3.$) and the biogas technology (in equilibria $4.$ and $5.$), could reach their maximum $\theta_h$ and $\theta_b$ and benefit from an unnecessary incentive. In particular, this occurs when $\theta_h$ or $\theta_b$ are respectively the minimum value in the $\min$ function appearing in Eqs. \eqref{NYC13}, \eqref{NYC14} and Eqs.~\eqref{NYC15}, \eqref{NYC16}, \eqref{NYC17}, \eqref{NYC18}. In this case, both the players push their investment to their upper limit $\theta_h$ and $\theta_b$. Since $\theta_h + \theta_b > d$, we see that the incentive $Z$ is excessive and favors speculative behavior of the players rather than an efficient ecological transition of their energy systems. 
}

\subsection{Upper-level problem: Stackelberg solution for the coordinator}

The coordinator pursues the viability of the energy community: to this end, it sets a Nash bargaining criterion to balance the economic advantage of both players in entering it, as introduced in the problem statement \eqref{NBarg}. Therefore, it seeks for the value $\beta^*$, which optimizes the following problem: 
\begin{eqnarray}
 \max_{\beta \in (0,1)} F(d,y_h^*, y_b^*,\beta), 
 \label{cooprob}
\end{eqnarray}

\noindent where $F$ is defined as 
$$ F(d,y_h^*, y_b^*,\beta) = (J_h(x_v,x_c,p,d,y_h^*, y_b^*,\beta) - d_h)(J_b(x_v,x_c,p,d,y_h^*, y_b^*,\beta) - d_b). $$
$d_h$ and $d_b$ are the disagreement points defined in \eqref{dishb} for the household and biogas producer, respectively. 

Notice again that in solving this problem  the coordinator acts strategically, considering $y_b^*$ and $y_b^*$, as the optimal responses of the two players to its decision, $\beta$. 

 Depending on the sign of $g_h$ and $g_b$, $F$ becomes
\begin{gather}
F(d,y_h^*, y_b^*,\beta) = \begin{cases}
    \beta(1- \beta) Z^2 w^2(\theta_h, \theta_b, d) & g_h \geq 0 , g_b \geq 0,\\
     \beta(1- \beta) Z^2 w^2(y_h^*, \theta_b, d) + (1 - \beta) Z w(y_h^* , \theta_b, d) g_h y_h^* & g_h < 0 , g_b \geq 0,\\
      \beta(1- \beta) Z^2 w^2(\theta_h, y_b^*, d) +  \beta Z w(\theta_h , y_b^*, d) g_b y_b^* & g_h \geq 0 , g_b < 0,\\
    \beta(1- \beta) Z^2w^2(y_h^*, y_b^*, d) +  \beta Z w(y_h^* , y_b^*, d) g_b y_b^* & \\
    +(1 - \beta)Z w(y_h^* , y_b^*, d) g_h y_h^* + y_b^* y_h^* g_b g_h & g_h \leq 0 , g_b \leq 0.
\end{cases}
\label{F_beta}
\end{gather}
We apply first order conditions to determine the maximizing $\beta$:
\begin{eqnarray*}
 \frac{\partial F(d,y_h^*, y_b^*,\beta)}{\partial \beta}  = 0
\end{eqnarray*}
The first case ($ g_h \geq 0 $ and $ g_b \geq 0 $) is the simplest, and the first order condition is solved trivially for $\beta = \frac{1}{2}$.

In the second and third cases of Eq. (\ref{F_beta}) the solution to the first order condition must be found numerically and it determines a unique solution $(y^*_h,y^*_b)$.

The fourth case ($g_h \leq 0$, $g_b \leq 0$) includes the relevant sub-case where $g_h \in (-Z \beta/(r + \lambda),0)$ and $ g_b \in  (-Z(1 - \beta))/(r + \lambda),0)$, which corresponds to Nash equilibrium 6. of Proposition \ref{ne} (the other sub-case, $g_h \le -Z \beta/(r + \lambda)$ and $g_b \le -Z(1 - \beta))/(r + \lambda)$, is irrelevant since neither players invest). 
Under this equilibrium, the coordinator can only fix $\beta=\beta_n$, otherwise no Nash equilibrium is possible. From the same proposition we know that, given $\beta_n$, there are infinitely many combinations $(y_h^*,y_b^*)$, which are all Nash equilibria and satisfy the conditions of Eqs. (\ref{NYC2}) and (\ref{NYC3}). 
Among them, the particular Nash equilibrium
\begin{eqnarray}\label{optinvbetan}
(y_h^*,y_b^*):y_h^*=y_b^*=\frac{y_h^*+y_b^*}{2},
\end{eqnarray}
where the sum $y_h^* + y_b^*$ is determined by Eqs. \eqref{NYC2} or \eqref{NYC3}, 
has the interesting property of maximizing $F$ for  $\beta = \beta_n$.
However, the fact that the two players will actually agree or not to set their investments according to (\ref{optinvbetan}) is not under the control of the coordinator. 

In practice, since all the combinations $(y_h^*,y_b^*)$ respecting either condition (\ref{NYC2}) or (\ref{NYC3}) are Nash equilibria, the first player to make the investment decision locks the decision of the other, as the second player will find it optimal to add an investment complementary to those conditions).

\section{Model application} \label{model_app}

\subsection{Data description and parameter estimation}




To estimate the parameters of the geometric Brownian motions describing the evolution of power demand, gas prices, and spot electricity prices, we consider hourly historical data of power demand, spot electricity price and gas price in Italy. The power demand, electricity price and gas price data are measured from January 2, 2014, to February 29, 2024. The power demand data corresponds to hourly measurements in MWh of aggregate power demand of Italy. The electricity spot price data correspond to hourly measurements in \euro/MWh of the Italian national price (PUN), while the gas spot price data corresponds to daily measures in \euro/MWh of the Italian PSV natural gas\footnote{both time series are available in the website of Gestore del Mercato Elettrico (GME), at {\tt https://www.mercatoelettrico.org/it/download/DatiStorici.aspx} for power prices and at {\tt https://www.mercatoelettrico.org/It/download/DatiStoriciGas.aspx} for gas prices}.

For the Brownian motion of the purchase electricity price, $X_c$, we set  the corresponding volatility, $\sigma_c$, equal to a small percentage of that of the spot electricity price. The reason is that, usually, the energy price applied by retailers to final consumers is an average of the spot price observed in a previous period, increased by a margin to cover costs and profit. Thus, by construction it is reasonable to set  $\sigma_c < \sigma_v$. 
We suppose that the volatility of the purchase price is approximately $1.47\%$ of the volatility of the spot electricity price.

To estimate the parameters of our three Brownian motions, we first remove seasonality from the three data series using harmonic regression \cite{ts}. 
The harmonic regression consists of finding the significant frequencies of a given time series sampled at discrete times $(X(t_i))_{i=1,\ldots,I}$
and representing it as a sum of trigonometric functions plus some noise. Suppose $f_k$, $k = 1, \ldots, n$ , are the significant frequencies, so for our data set, we approximate the seasonality as

\begin{eqnarray}\label{hr}
    X(t_i) = \sum_{k = 1}^n A_k \cos(2 \pi f_k t_i) + B_k \sin(2 \pi f_k t_i) + \epsilon(t_i),
\end{eqnarray}

\noindent where $\epsilon$ is some noise. For the power demand, we suppose that the data can be described by \eqref{hr}, where the noise $\epsilon$ is described by a geometric Brownian motion. On the other hand, for the electricity and gas spot prices, we suppose that the logarithm of the observations can be described by \eqref{hr}. In these cases, the noise $\epsilon$ is described by Brownian motions with drift. 
}
The significant frequencies for the three data series and the corresponding amplitudes are presented in Table \ref{fd} in Appendix A.



We estimate the parameters $\sigma_d$, $\sigma_v$ and $\sigma_p$ of the three geometric Brownian motions associated with the power demand, electricity price and gas price by using least squares estimator as in \cite{brigo}, see Table \ref{ex1_ini}. 
In principle, also the drifts $\mu_d$, $\mu_v$, $\mu_p$ can be estimated with the same technique. However, 
a quite natural scenario used to set up a REC is where the members suppose that the "world will proceed as it is" in that moment. For them, agreeing on whether trends or volatilities will definitely increasing or decreasing in the long run is of paramount difficulty. 
Indeed, fixing the plant capacity and the sharing rules are decisions that tend to remain fixed over time for quite obvious reasons. 
Thus, a world evolving in a martingale way is a neutral and natural setting which is interesting to adopt for our application. 
As our main purpose is to model an energy community and not to model prices or power demand, we manipulate the values for the three drifts $\mu_d$, $\mu_v$ and $\mu_p$ in order to obtain martingale processes for $D$, $X_v$ and $P$, while maintaining the values of the volatilities. We report the values of these modified drifts in the column "Martingale values" of Table \ref{ex1_ini}, which are obtained by letting $\tilde \mu_x := - \sigma^2_x/2$ for $x = c, d, p, v$. 
A natural consequence is that the quantities $r_x$, $x = c, d, p, v$, defined in Assumption 3.1, are all equal to $r$. 
Since all the quantities to be optimized depend on $\mu_x$ and $\sigma_x$, for $x = c, d, p, v$, only through the corresponding $r_x$, any change in $\sigma_x$ compensated by a corresponding change in $\mu_x$ would still result in $r_x = r$, without changing the corresponding equilibria. Instead, a change in demand volatility $\sigma_d$, even if compensated by a corresponding change in $\mu_d$, would affect $m_{1,2}$, and thus the coefficients $B_1$ and $C_1$ in the analytic expression for the unitary incentive $w$. For this reason, the only sensitivity worth studying for the geometric Brownian motions is with respect to $\sigma_d$. 



{
\begin{table}
\centering
\begin{tabular}{cclccclcc} 
\hline
    & Initial &  & Drift & Estim.~      & Martingale~ &  & Diffusion & Estim.~   \\
condit. & Value  &  & parameter     & value  & value  &  & parameter     & value  \\ 
\cline{1-2}\cline{4-6}\cline{8-9}
$x_v$     & $80$ \euro/MWh &  & $\mu_v$ (1/h) & $2.824 \cdot 10^{-6}$   & $-0.0052$            &  & $\sigma_v$ (1/$\sqrt{\mathrm h}$) & 0.102199   \\
$x_c$     & $85$ \euro/MWh &  & $\mu_c$ (1/h) & —                     & $-2.140\cdot 10^{-6}$  &  & $\sigma_c$ (1/$\sqrt{\mathrm h}$) & 0.0015   \\
$p$       & $45$ \euro/MWh &  & $\mu_p$ (1/h) & $- 6.313 \cdot 10^{-7}$ & $-1.099\cdot 10^{-4} $ &  & $\sigma_p$ (1/$\sqrt{\mathrm h}$) & $0.01482$  \\
$d$       & $0.30$ MW  &  & $\mu_d$ (1/h) & $1.017 \cdot 10^{-6}$   & $-7.898 \cdot 10^{-4}$ &  & $\sigma_d$ (1/$\sqrt{\mathrm h}$) & 0.0397     \\
\hline
\end{tabular}
\caption{Initial condition, drift and volatilities for the purchase and sale spot electricity prices, spot gas price and power demand. Recall that the market data do not cover $X_c$, whose volatility $\sigma_c$ is fixed as $\sigma_c := 0.015 \sigma_v$.}
\label{ex1_ini}
\end{table}
}

\subsection{Optimal solution}

\label{example1} 
Table \ref{ex1_ctte} presents the parameter values used in the application of our model.
\begin{table}[H]
    \centering
    \begin{tabular}{|c| c |}
    \hline
         Parameter &  Value \\
          \hline
         $r$ & $3.4247 \cdot 10^{-6}$ 1/h \\
          \hline
         $c_h$ &   $2500000$ \euro/MW \\
          \hline
         $c_b$ & $900000$ \euro/MW \\
          \hline
         $\theta_h$ & $0.32$ MW\\
          \hline
         $\theta_b$ & $0.20$ MW\\
          \hline
         $Z$ & $110$ \euro/MWh\\
          \hline
         $K_g$ & $18.9394$ $m^3$\\
          \hline
          $\rho_c$ & 0.01 \\
           \hline
         $\lambda$ & 0.00001\\
          \hline
          $\alpha$ & 1500/8760\\
          \hline 
          $\gamma$ & 0.6\\
          \hline
          $b$ & 0.01056 MWh/$m^3$\\
          \hline
    \end{tabular}
    \caption{Reference values of the parameters used in the application}
    \label{ex1_ctte}
\end{table}

A few comments on the values in Table \ref{ex1_ctte} follow. The value for $r$ corresponds to an annual discount rate of $3 \%$. The values for $c_h$ and $c_b$ are average (unit) investment costs commonly observed for PV and small size gas-turbine generation plants.
In general, the upper bounds $\theta_h$ and $\theta_b$ can be determined either by technical (e.g., available roof surface for the PV plants) or economic (e.g., budget) reasons, whichever applies first. 
In our application, we consider a maximum capacity of the biogas producer equal to  $\theta_b=0.2$ MW. Given a conversion factor $b = 0.01056$ MWh/m$^3$ \cite{GastEle}, this corresponds to using all the biogas produced in one hour ($K_g = 18.9394$ m$^3$) to reach the maximum output capacity $\theta_b$. 
We also assume that the household's investment capacity is bounded only by a budget constraint of \EUR $ 800'000$; thus, considering a PV installation costs of $  c_h = 2'500'000$\EUR/MW, we have a maximum PV capacity $\theta_h = 0.32$MW. 
The value $Z=110$ \euro/MWh\ is the example of the Italian incentive tariff for RECs, which represents the first known application of the "virtual" framework. 
Considering a simplified constant  solar radiation of about 4.10 h per day, or equivalently $1500$ operational hours per year, we obtain a value of $\alpha =0.1712$, which corresponds to the usual ratio of the operative hours of a PV panel in a year. For the gas-to-power (G2P) efficiency factor $\gamma$, we consider the value for gas-fired power plants, which is approximately $0.6$.  
Finally, we fix a loosely positive  correlation parameter $\rho_c = 0.01$ between the purchase electricity spot price and the power demand. Indeed, it is well known that electricity markets are rather inelastic and demand reacts very slowly to changes in prices.

\begin{table}[H]
    \centering
    \begin{tabular}{|c|c|}
    \hline
       Parameter  & value \\
       \hline
       $\beta_h$ & $-$11.3172\\
       \hline
         $\beta_b$ & 0.8189\\
         \hline
         $g_h$ & $1.5\cdot 10^{6}$\\
         \hline
         $g_b$ & $-2.4 \cdot 10^{4}$\\
         \hline
    \end{tabular}
    \caption{Parameters of the realistic example}
    \label{ex1_case}
\end{table}

\noindent The values in Table \ref{ex1_case}, which result from those in Table (\ref{ex1_ctte}), are key to assessing the kind of equilibrium. We see that our case corresponds to equilibrium 4.1 ($g_h > 0$ and $g_b < 0$) in Equation \eqref{NYC15}, Proposition \ref{ne}. Under this type of equilibrium, the household does not need any incentive to invest in PV panels. However, the biogas producer invests (and enters the REC) only if  
$$g_b + Z \frac{(1 - \beta)}{(r + \lambda)} >0.$$ 
This is a case where the coordinator, by fixing the value of $\beta^*$ according to problem (\ref{cooprob}), determines both the birth of the REC and a fair sharing rule for the incentives among the two members. Table \ref{sol_real} shows the optimal installations for the members $(y_h^*,y_b^*)$ and the fair sharing rule $\beta^*$.

\begin{figure}[H]
    \centering
    \includegraphics[scale = 0.3]{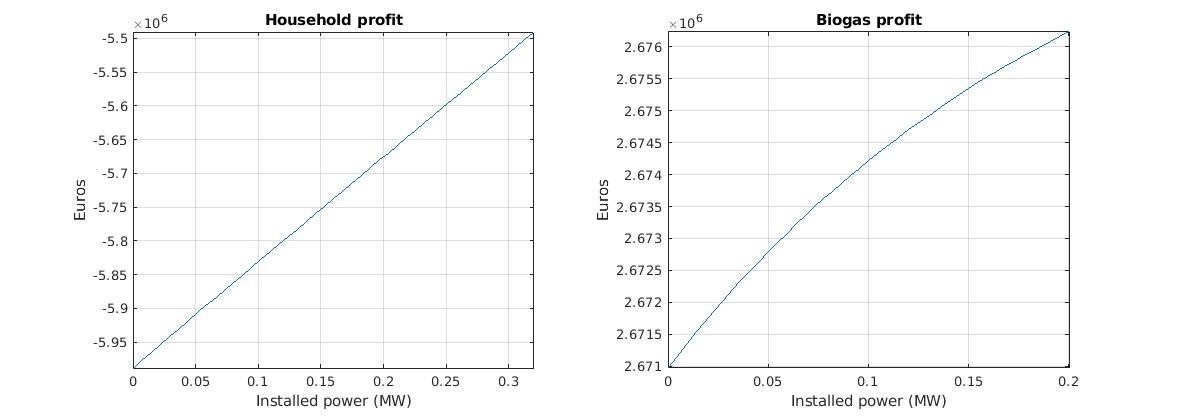}
    \caption{Household and biogas producer best response profits}
    \label{ex_1_plot}
\end{figure}

\begin{table}[H]
    \centering
    \begin{tabular}{|c|c|}
    \hline
       Parameter  & value \\
       \hline
         $y_h^*$ & $0.32$ MW\\
         \hline
         $y_b^*$ & $0.2$ MW\\
         \hline
       $\beta^*$ & 0.4765\\
       \hline
    \end{tabular}
    \caption{Optimal solution for the example}
    \label{sol_real}
\end{table}
\noindent Figure \ref{ex_1_plot} shows the profits for the household (left) and the biogas producer (right), considering the solution value of $\beta^*=0.4765$. Both profit plots are calculated strategically, that is, including the optimal investment choice of the other member. 

We see that both profit functions increase monotonically, so the optimal capacity investment for the two members coincides with the corresponding maximum capacity boundaries. The resulting total generation capacity of this REC ($0.52$ MW) is significantly higher than the average internal energy demand, i.e., $d=0.3$ MW. We observe that, in this case, the self-consumption incentive favors the birth of the REC, as the biogas producer would not have installed any capacity without its presence. However, at the same time, the amount fixed for the incentive generates a speculative investment on the biogas producer's side. Finally, this incentive is unnecessary for the household, as the corresponding installation would have been the maximum one even without the presence of incentives.

\subsection{Sensitivity analysis}

We conduct a sensitivity analysis of the optimal solution with respect to the parameters $x_v$, $p$, $d$, $r$, $c_h$, $c_b$, and $\sigma_d$. Using the parameter values established in Section 5 as a reference, we analyze each parameter individually by varying its value while keeping the others constant. Through graphical plots, we illustrate the effects of these parameters on investment decisions ($y^*_h, y^*_b$), profits ($J_h, J_b$), the fair sharing rule ($\beta$), and the joint surplus ($F$).

Our focus is on the parameter intervals where significant impacts on investment decisions are observed. In scenarios where the community is not formed, we set $\beta^* = 0$ as a convention for graphical representation. In such cases, when the sharing rule is depicted as zero, it indicates that the incentive mechanism is not considered.


We can partition the parameters into two categories, depending on the type of impact they generate:
\begin{itemize}
\item "All or nothing" impact: 
these parameters exhibit a binary effect, where their range primarily results in either maximum or zero investment from both parties. A narrow critical range exists within the parameter's interval, dividing it into two distinct outcomes: one where both parties make a maximum investment and another where no investment is made. Within this critical range, investment decisions may fall between these extremes and are not strictly zero or maximum.
\item  Incremental impact: these parameters have a gradual effect on optimal investment decisions as they vary over a substantial range. Their influence enables the coordinator to address the problem effectively, thereby fostering the establishment of the REC.
\end{itemize}
Table \ref{sum_sen} summarizes this classification of the parameters.

\begin{table}[H]
    \centering
    \begin{tabular}{| c| c | }
    \hline
       All or nothing   & Incremental  \\
       \hline
       $x_v$, $p$
       &  $c_b$, $c_h$, $r$, $\sigma_d$, $d$ \\
\hline
    \end{tabular}
    \caption{Summary of parameter sensitivity}
    \label{sum_sen}
\end{table}


\subsubsection{"All or nothing" impact}


Let us start by presenting the results for the "all or nothing" parameters. 

\begin{table}[H]
    \centering
    \begin{tabular}{|c|c|c|}
    \hline
       $x_v$ value  &  Optimal investment decision & Sharing rule\\
        \hline
         $[0, 50.01)$ & $y_h^* = 0$, $y_b^* = 0$ & $\beta^* = 0$  \\
        \hline
         $[50.01, 79.733)$ &  $y_h^* = \theta_h$, $y_b^* = 0$ & $\beta^* = 0$  \\
         \hline 
         $[79.733,80.137)$ & $y_h^* = \theta_h$, $y_b^* = \min \left\{ d \left( \frac{1 - \beta}{1 - \beta_b} \right)^{1/m_2} - \theta_h, \theta_b \right\}$ & $\beta^* \in [0.4, 0.5) $ \\
         \hline 
         $[80.137, +\infty)$ & $y_h^* = \theta_h$, $y_b^* = \theta_b$ & $\beta^* = 0.5$ \\
         \hline
    \end{tabular}
    \caption{Sensitivity w.r.t. electricity price $x_v$}
    \label{sen_thre_x_v}
\end{table}

Table \ref{sen_thre_x_v} shows the electricity price threshold values, which trigger the biogas producer's and the household's maximum investments. For electricity prices lower than $x_v = 50.01$ \euro /MWh, both players deem it optimal not to install. This changes at $x_v = 50.01$ \euro /MWh, when, for the biogas producer it is still optimal not to install, while for the household it becomes optimal to install its full capacity $\theta_h$ as $g_h$ becomes strictly positive. Going on, the investment of the biogas producer becomes profitable from $x_v = 79.733$ \euro /MWh, thanks to the incentives, corresponding to Equilibrium 4.1; however, from $x_v = 80.137$ \euro/ MW onward, the equilibrium changes to Equilibrium 1., as in this case the investment is profitable even without a monetary incentive. In summary, from electricity prices higher than $x_v = 79.733$ \euro /MWh, the REC is created.
 
\begin{table}[H]
    \centering
    \begin{tabular}{|c|c|c|}
    \hline
       $p$ value  &  Optimal investment decision & Sharing rule\\
        \hline
         $[0, 44.918)$ & $y_h^* = \theta_h$, $y_b^* = \theta_b$ & $\beta^* = 0.5$  \\
         \hline 
         $[44.918, 45.161)$ & $y_h^* = \theta_h$, $y_b^* = \min \left\{ d \left( \frac{1 - \beta}{1 - \beta_b} \right)^{1/m_2} - \theta_h, \theta_b \right\}$ & $\beta^* \in [0.39, 0.5)$ \\
         \hline 
         $[45.161, + \infty)$ & $y_h^* = \theta_h$, $y_b^* = 0$ & $\beta^* = 0$ \\
         \hline
    \end{tabular}
    \caption{Sensitivity w.r.t. gas price $p$}
    \label{sen_thre_p}
\end{table}

Table \ref{sen_thre_p} shows the change in the decision of the biogas producer with respect to changes in $p$. In contrast to the electricity price case, the threshold value triggers only the no-investment decision of the biogas producer. The optimal installation is equal to the maximum possible $\theta_b$ for gas prices lower than $p = 44.918$ \euro/MWh. From that threshold onward, it is still profitable for the biogas producer to stay in the REC. However, this changes rapidly, as from gas prices higher than $p = 45.161$ \euro/MWh the optimal installation for the biogas producer becomes zero, and the REC is no longer viable.

\subsubsection{Sensitivity with respect to the turbine cost $c_b$}  

In this and the following subsections we conduct a sensitivity analysis with respect to the parameters $c_b$, $c_h$, $r$, $\sigma_d$ and $d$. The discussion is based on four figures, showing how:
\begin{itemize}
    \item the optimal installed capacities $(y_h,y_b)$,
    \item the expected profits $(J_h,J_b)$,
    \item the optimal sharing rule, $\beta$, and the discounted incentive $w$,
    \item the coordinator objective,
\end{itemize}
change depending on the given parameter.
\vspace{10pt}

\begin{figure}[H]
\centering
     \begin{subfigure}[b]{0.45\textwidth}
         \centering
         \includegraphics[width=\textwidth]{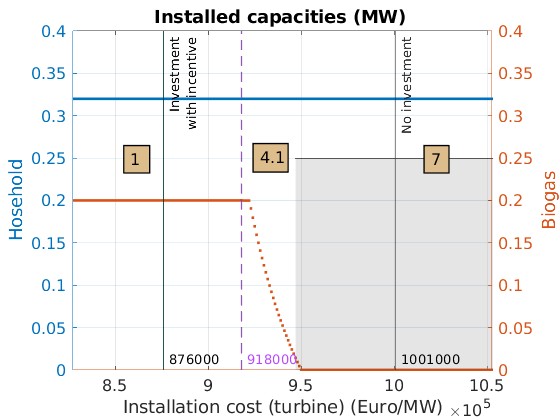}
         \caption{Optimal capacities for different values of $c_b$}
         \label{ex_1_br_cb}
     \end{subfigure}
    \hfill
    \begin{subfigure}[b]{0.45\textwidth}
         \centering
         \includegraphics[width=\textwidth]{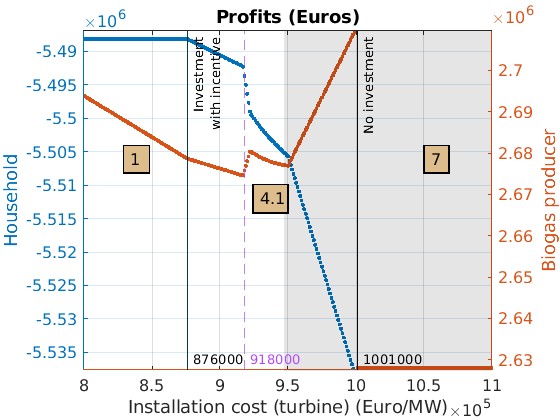}
         \caption{Profits for different values of $c_b$}
         \label{ex_1_J_cb}
    \end{subfigure}
\caption{Sensitivity w.r.t. $c_b$}
\label{binst_cb}
\end{figure}

\begin{figure}[H]
\centering
  \begin{subfigure}[t]{0.45\textwidth}
         \centering
         \includegraphics[width=\textwidth]{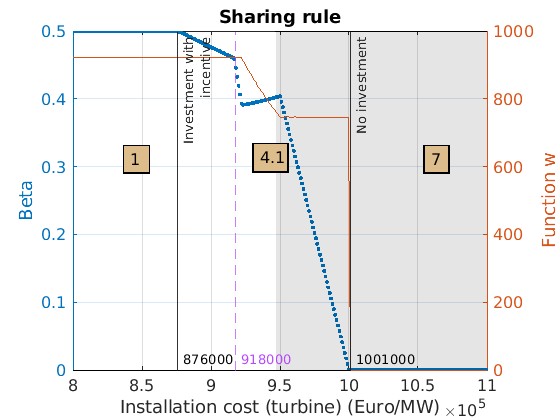}
         \caption{Optimal $\beta$ for different values of $c_b$}
         \label{ex_1_b_c_b}
     \end{subfigure}
    \hfill
     \begin{subfigure}[t]{0.45\textwidth}
         \centering
         \includegraphics[width=\textwidth]{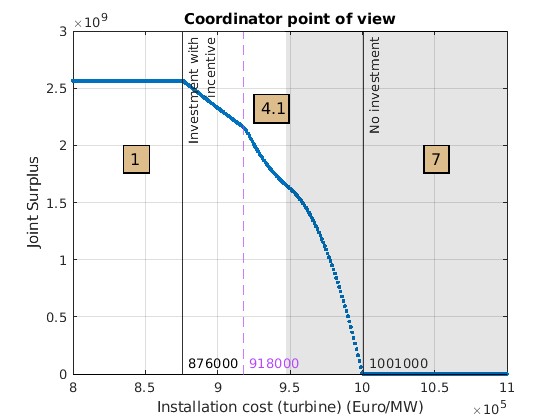}
         \caption{Coordinator objective function $F$ for different values of $c_b$}
         \label{ex_1_F_cb}
     \end{subfigure}
    \caption{Sensitivity w.r.t. $c_b$}
    \label{b_F_cb}
\end{figure}

We begin by focusing on the turbine cost parameter, $c_b$. As expected, a higher cost for the G2P turbine reduces the investment of the biogas producer. 
This anticipated outcome is confirmed by our results, as shown in both panels of Figure \ref{ex_1_br_cb}. However, Figures \ref{ex_1_J_cb} and \ref{ex_1_F_cb} reveal insightful details, highlighting the complex interdependence among the decision variables in our model.

A first thing to observe is that three different equilibria occur as the investment cost of the turbine increases, namely equilibrium 1, 4.1 and 7 (respectively in eqs. \eqref{NYC11},  \eqref{NYC15} and  \eqref{NE7} of Proposition \ref{ne}). 
According to equilibrium 1, 
both REC members invest up to $\theta_h$ and $\theta_b$ as long as the turbine cost $c_b$ remains below 876'000 \euro/MW. In this case, although 
no incentive is required for both players to invest, the REC is still created, since it provides an additional incentive for both players. Under this condition, the coordinator optimally sets $\beta^* =0.5$ (see Figure \ref{ex_1_b_c_b}).

If the turbine cost falls within the range [876'000 \euro/MW, 1'001'000 \euro/MW], the new equilibrium is 4.1:
the coordinator provides a higher incentive to the biogas producer to compensate for the higher turbine cost (see Figure \ref{ex_1_b_c_b} and recall that a lower $\beta$ means a higher incentive assigned to the biogas producer). 
As a result, the biogas producer initially maintains the same investment level at $\theta_b$. However, if the turbine cost exceeds 918'000 \euro/MW, the biogas producer, despite the  increased incentive share, begins to reduce its investment (see Fig. \ref{ex_1_br_cb} and $\beta$ in Fig. \ref{ex_1_b_c_b} for $c_b \ge$ 918'000 \euro/MW). 
The investment decrease is abrupt, and soon it reaches the grey area in Fig. \ref{ex_1_br_cb}, which corresponds to an investment level technically meaningless ($<$ 1KWh), which is non-zero only mathematically. From this point onward, no REC can be realistically created. 
When the turbine cost $c_b \ge$ 1'001'000 \euro/MW, we enter equilibrium 7, and the biogas investment is zero, even mathematically. 

As we can see, the significant reduction in the turbine capacity (occurring if $c_b \ge$ 918'000 \euro/MW), triggers nonlinear effects to the function $w$  (i.e. the expected cumulative self-consumption from time $0$ to $\tau$) and to $\beta$ (Fig. \ref{ex_1_b_c_b}). 
After a short and rapid effort of the coordinator to sustain $w$ by  reducing $\beta$ below 0.4, the sharp drop in the biogas investment becomes unavoidable and  rapidly reduces $w$ from about 950 down to about 750. 
This forces the coordinator to adjust  its policy again. Indeed, to keep a balance between the profit of the two REC members, the coordinator begins to gradually increase $\beta$. 
As we see in Figs. \ref{ex_1_J_cb} and \ref{ex_1_F_cb}, the profits of the two players decrease as $c_b$ increases, with only a brief but intense interruption where the coordinator "rushes" to sustain $w$ in favor of the biogas producer. 
While this "intervention" breaks the profit pattern for the two players, we see in Fig. \ref{ex_1_F_cb}, that from its point of view, the coordinator follows a smooth pattern, favoring the biogas producer to compensate it for the increasing turbine cost. 
No further comments are necessary from the grey area on, since solutions have no practical meaning. 

\subsubsection{Sensitivity with respect to the PV panel cost $c_h$} 

\begin{figure}[H]
\centering
  \begin{subfigure}[b]{0.45\textwidth}
         \centering
         \includegraphics[width=\textwidth]{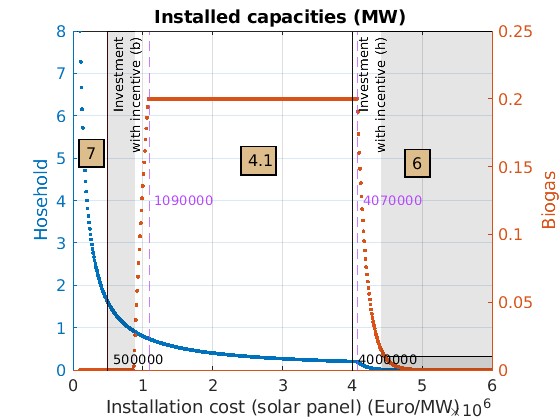}
         \caption{Optimal capacities for different values of $c_h$}
         \label{ex_1_br_ch}
     \end{subfigure}
     \hfill
     \begin{subfigure}[b]{0.45\textwidth}
         \centering
         \includegraphics[width=\textwidth]{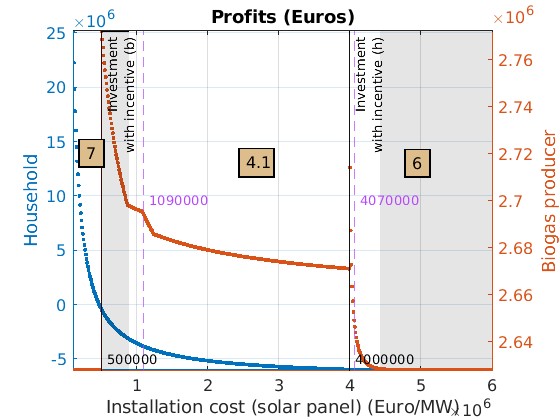}
         \caption{Profits for different values of $c_h$}
         \label{ex_1_J_ch}
     \end{subfigure}
        \caption{Sensitivity w.r.t. $c_h$}
        \label{S_ch}
\end{figure}

The following analysis provides an interesting example of how the decision of one member affects the other's decision. 
As we can see in Figure \ref{ex_1_br_ch}, increasing PV panels cost monotonically reduces  the total installation of the household. This is easily explained by considering the budget constraint, as $\theta_h = \mbox{budget} / c_h$. Higher values of cost $c_h$ reduce $\theta_h$, as the household maintains the budget fixed.

\begin{figure}[H]
\centering
  \begin{subfigure}[b]{0.45\textwidth}
         \centering
         \includegraphics[width=\textwidth]{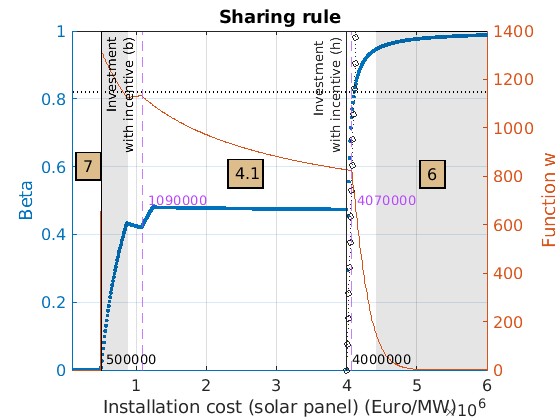}
         \caption{Optimal $\beta$ for different values of $c_h$}
         \label{ex_1_b_c_h}
     \end{subfigure}
    \hfill
     \begin{subfigure}[b]{0.45\textwidth}
         \centering
         \includegraphics[width=\textwidth]{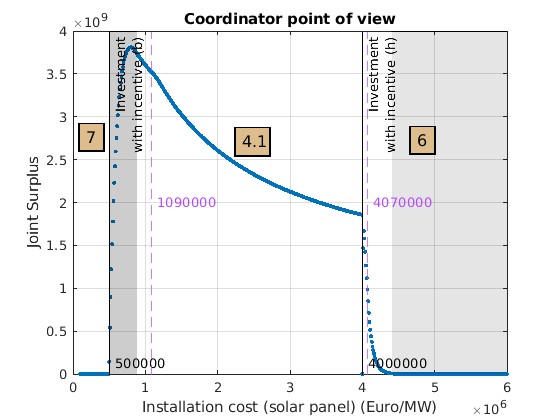}
         \caption{$F$ for different values of $c_h$}
         \label{ex_1_F_ch}
     \end{subfigure}
        \caption{Sensitivity w.r.t. $c_h$}
        \label{b_F_ch}
\end{figure}

At the beginning, for installation costs $c_h$ below approximately $1'090'000$ \EUR/MW, the system operates under equilibrium type 7, as indicated in Figures \eqref{S_ch} and \eqref{b_F_ch}. In this range, the household's installed capacity decreases rapidly as costs increase, while the biogas producer installation capacity remains equal to $0$. 

In the intermediate range, the system transitions to equilibrium type 4.1, where the sharing rule \( \beta \) initially rises steeply from zero to approximately \( 0.45 \). Subsequently, within the cost interval from  $1'090'000$ \EUR/MW to around $4'000'000$ \EUR/MW, \( \beta \) stabilizes at approximately \( 0.48 \), ensuring that the biogas producer receives an adequate incentive to maintain full investment capacity.
As highlighted in Figure \ref{ex_1_br_ch}, also the household continues to invest, albeit at a reduced rate compared to equilibrium 7. Under this sharing rule, both members install a total capacity significantly higher than the average demand $d$.
However, the orange line representing the function \( w \)  gradually declines, indicating a reduction in self-consumption levels. This decline occurs as random demand peaks are progressively lost due to the decreasing PV capacity. 

Finally, for installation costs exceeding $4'000'000$ \EUR/MW, the system shifts to equilibrium type 6, where a significant change occurs. The sharing rule \( \beta \) transitions to \( \beta = \beta_n \), instead of the value that would maximize the joint surplus \eqref{cooprob}. The household's incentive requirement increases. Yet, higher levels of $\beta$ do not prevent the sharp decline in its installed capacity, while the biogas producer continues at maximum capacity until $4'070'000$ \EUR/MW. Beyond this point, both players begin to reduce their investments rapidly. 
The condition \( \beta^* > \beta_b \), which occurs when \( c_h > \) $4'070'000$ \EUR/MW, triggers a transition to Nash equilibrium type 6.2. In this equilibrium, the total REC capacity \( y_h + y_b \) falls below the average demand \( d \), marking a significant shift in the system's operational balance.

\subsubsection{Sensitivity with respect to the risk free rate $r$} This is another example of how both members compete for power installation. 

\begin{figure}[H]
\centering
  \begin{subfigure}[b]{0.45\textwidth}
         \centering
         \includegraphics[width=\textwidth]{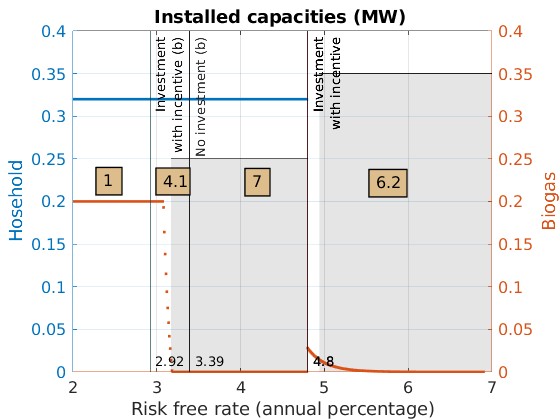}
         \caption{Optimal capacities for different values of $r$}
         \label{ex_1_br_r}
     \end{subfigure}
     \hfill
     \begin{subfigure}[b]{0.45\textwidth}
         \centering
         \includegraphics[width=\textwidth]{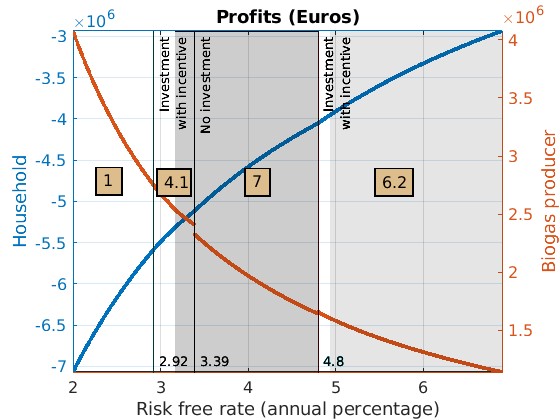}
         \caption{Profits for different values of $r$}
         \label{ex_1_J_r}
     \end{subfigure}
        \caption{Sensitivity w.r.t. $r$}
        \label{S_r}
\end{figure}

Figure \ref{ex_1_br_r} illustrates the changes in investment decisions for different values of \( r \). 
For values of \( r \) between \( 2\% \) and \( 2.92\% \), both members find it beneficial to invest regardless of the incentive, corresponding to Equilibrium 1 (Eq. \eqref{NYC11}). 

\begin{figure}[H]
\centering
  \begin{subfigure}[b]{0.45\textwidth}
         \centering
         \includegraphics[width=\textwidth]{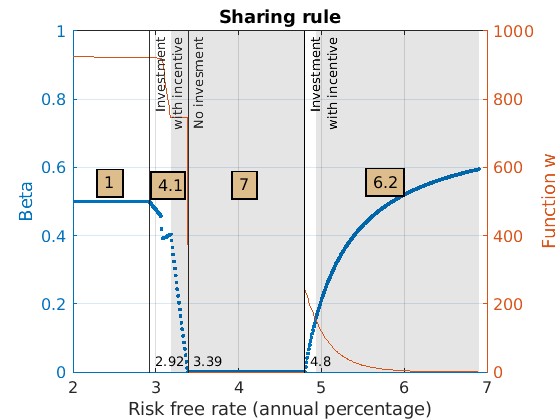}
         \caption{Optimal $\beta$ for different values of $r$}
         \label{ex_1_b_r}
     \end{subfigure}
     \hfill
     \begin{subfigure}[b]{0.45\textwidth}
         \centering
         \includegraphics[width=\textwidth]{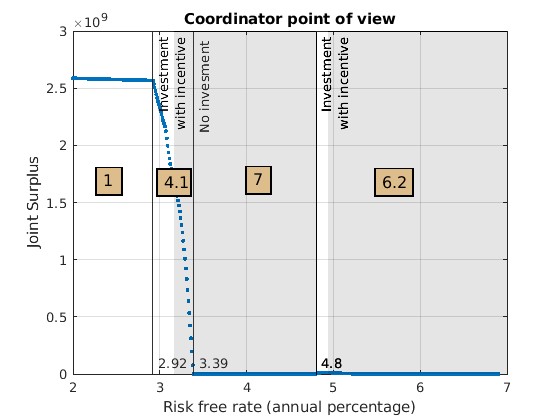}
         \caption{$F$ for different values of $r$}
         \label{ex_1_F_r}
     \end{subfigure}
        \caption{Sensitivity w.r.t. $r$}
        \label{b_F_r}
\end{figure}

Between \( 2.92\% \) and \( 3.39\% \), the biogas producer requires an incentive to invest, leading to Equilibrium 4.1 (Eq. \eqref{NYC15}). As shown in the figures, the incentive is initially sufficient to encourage the biogas producer to invest at its maximum capacity; however, it decreases rapidly until reaching \( 3.93\% \), where the biogas producer ceases investment and the REC is no longer created. 
In Figure \ref{ex_1_b_r}, we observe that for values of \( r \) up to \( 3.39\% \), the coordinator significantly reduces \( \beta \) to nearly zero. This allocation strategy ensures that the majority of the incentive is directed toward the biogas producer, encouraging continued investment in G2P capacity.

At \( r = 4.8\% \), the household requires an incentive to invest in PV panels, leading to the transition to Equilibrium 6.2 (Eq. \eqref{NYC3}). At this point, the PV power installation is sufficiently low, allowing the biogas producer to re-enter the REC (see Fig. \ref{ex_1_br_r}) with the help of the incentive, though at capacity levels well below the expected energy demand of the community. According to equilibrium type 6.2, the sharing rule \( \beta^* \) that ensures an equilibrium is \( \beta_n \). However, for risk-free rates exceeding \( 5\% \), power installation becomes technically infeasible (Figure \ref{ex_1_b_r}).

\subsubsection{Sensitivity with respect to the power demand volatility $\sigma_d$} 

\begin{figure}[H]
\centering
  \begin{subfigure}[b]{0.45\textwidth}
         \includegraphics[width=\textwidth]{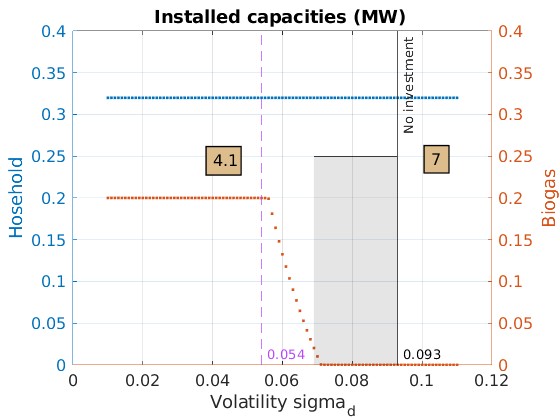}
         \caption{Optimal capacities for different values of $\sigma_d$}
         \label{ex_1_br_sigma_d}
     \end{subfigure}
    \begin{subfigure}[b]{0.45\textwidth}
         \centering
         \includegraphics[width=\textwidth]{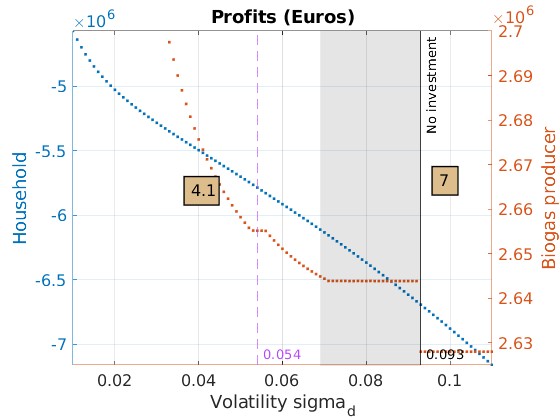}
         \caption{Profits for different values of $\sigma_d$}
         \label{ex_1_J_sigma_d}
     \end{subfigure}
     \vfill
 \begin{subfigure}[b]{0.45\textwidth}
         \centering
         \includegraphics[width=\textwidth]{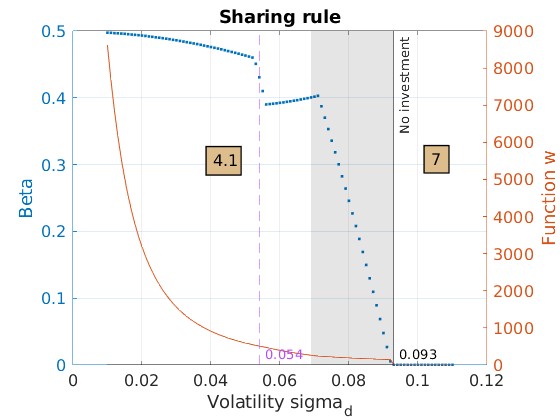}
         \caption{Optimal $\beta$ for different values of $\sigma_d$}
         \label{ex_1_b_sigma_d}
     \end{subfigure}
     \begin{subfigure}[b]{0.45\textwidth}
         \centering
         \includegraphics[width=\textwidth]{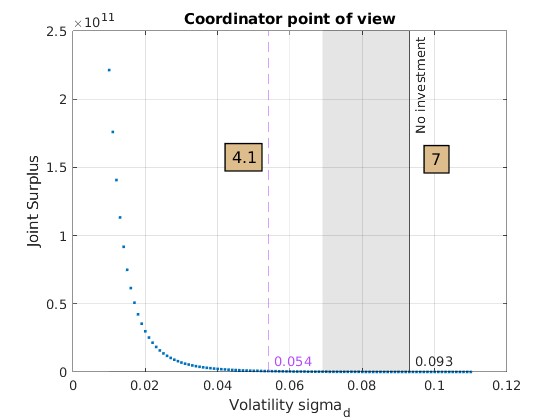}
         \caption{$F$ for different values of $\sigma_d$}
         \label{ex_1_F_sigma_d}
     \end{subfigure}
        \label{b_F_sigma_d}
     
     \caption{Sensitivity w.r.t. $\sigma_d$}
        \label{S_sigma_d}
\end{figure}

The four plots in Figure \ref{S_sigma_d} illustrate the sensitivity of the optimal solution with respect to 
the energy demand's volatility \(\sigma_d\). Notably, in this case, the demand drift \(\mu_d\) is adjusted simultaneously with \(\sigma_d\) to preserve the martingale property of the stochastic process \(D_t\), ensuring that the expected demand remains constant, i.e., \(\mathbb{E}(D_t) = d\).

A key explanation for the following results is that the function defining the self-consumption in Eq. \eqref{selfc}, i.e., $\min(D^d(t), y_h + y_b)$), is concave in demand. As a consequence, increasing \(\sigma_d\) while keeping the expected demand \(d\) constant reduces the expected value of the incentive. This explains the systematic drop in the profits of both members, especially for the biogas producer (Figure \ref{ex_1_J_sigma_d}). The critical volatility threshold at about $\sigma_d=0.093$ (highlighted in all the four plots) marks the shift from Equilibrium 4.1, where biogas invests thanks to the incentive, to Equilibrium 7, where only the household continues investing.

In Figure \ref{ex_1_br_sigma_d}, the household maintains a stable installation capacity regardless of \(\sigma_d\), reflecting its independence from the incentive. However, the biogas producer reacts strongly to the increase in demand volatility, reducing G2P investment when \(\sigma_d\) exceeds the threshold of 0.054. This adjustment allows the biogas producer to recover some profitability by shifting its strategy toward selling biomethane rather than generating electricity. This effect is visible in Figure \ref{ex_1_J_sigma_d}, where biogas profits experience a slight recovery precisely at \(\sigma_d \approx 0.054\), despite the overall declining trend. 

The impact of this strategic shift is also apparent in the sharing rule \(\beta\) shown in Figure \ref{ex_1_b_sigma_d}. In the interval \(0.056 \leq \sigma_d \leq 0.07\), there is a sudden and temporary increase in \(\beta\), reflecting the redistribution of incentives to accommodate the altered profit imbalance between the household and the biogas producer. However, as \(\sigma_d\) continues to rise beyond this interval, biogas investment collapses completely, leading to the eventual disappearance of any meaningful sharing dynamics.

Finally, from the coordinator’s perspective (Figure \ref{ex_1_F_sigma_d}), the joint surplus \(F\) decreases with increasing \(\sigma_d\), confirming the detrimental effect of high volatility on the overall economic viability of the REC. Once \(\sigma_d\) surpasses 0.093, the REC ceases to exist, as investment incentives are no longer sufficient to sustain participation.

These results highlight two fundamental insights: first, the negative impact of demand volatility on the expected self-consumption incentive, which disproportionately affects the biogas producer due to its reliance on electricity sales. Second, the biogas producer's ability to mitigate losses by dynamically shifting investment strategies, temporarily stabilizing profitability before an eventual decline. 

\subsubsection{Sensitivity with respect to the power demand $d$}

\begin{figure}[H]
\centering
  \begin{subfigure}[b]{0.45\textwidth}
         \centering
         \includegraphics[width=\textwidth]{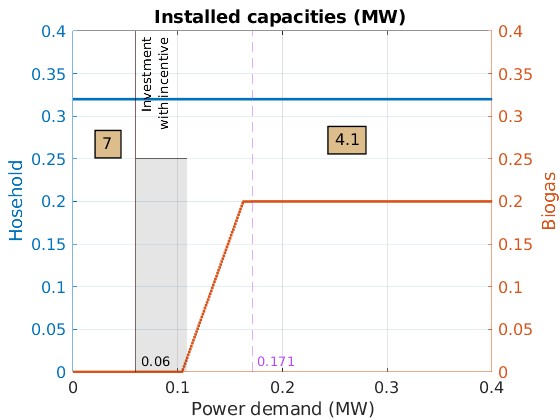}
         \caption{Optimal capacities for different values of $d$}
         \label{ex_1_br_d}
     \end{subfigure}
     \hfill
     \begin{subfigure}[b]{0.45\textwidth}
         \centering
         \includegraphics[width=\textwidth]{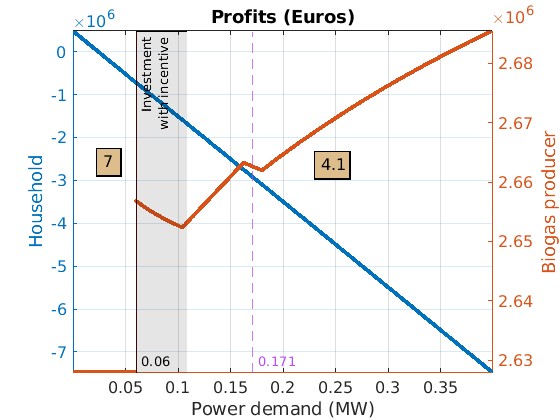}
         \caption{Profits for different values of $d$}
         \label{ex_1_J_d}
     \end{subfigure}
        \caption{Sensitivity w.r.t. $d$}
        \label{S_d}
\end{figure}

The sensitivity analysis with respect to the average power demand \( d \) inside the REC highlights its significant influence on the biogas producer’s investment decision, similar to the effect observed with power demand volatility. 
As shown in Figure \ref{ex_1_br_d}, the household consistently installs at maximum capacity, independent of the value of \( d \), while the biogas adjusts its investment level depending on $d$. 
Specifically, starting from a threshold demand level, approximately \( d = 0.06 \) MW, the equilibrium type changes from 7 to 4.1. as the incentive becomes   minimally significant for the biogas producer. Below this point, the economic viability of investment remains insufficient, preventing the biogas producer from participating in the REC. However, the biogas producer requires a demand level of about $0.11$ MWh to utilize its G2P capacity above $0.02$ MW. 

The economic reasoning behind this behavior is linked to the profit and cost structures of the two REC members. As the household is already at its maximum installation $\theta_h$, it cannot expand further its PV capacity, so it experiences a decline in profitability as \( d \) increases. In contrast, as shown in Figure \ref{ex_1_J_d}, the biogas producer, thanks to revenues generated from selling electricity and benefiting from the incentive \( Z \), sees increasing profits with higher \( d \). This is due to the fact that larger shared energy volumes lead to an increase in  \( w \), strengthening the biogas producer’s revenue stream.

Notably, at \( d = 0.171 \) MW, both members reach their maximum installed capacity limits. However, due to the differing economic incentives, the sharing rule does not naturally settle at an equal split (\(\beta = 0.5\)). Instead, the optimal sharing rule is determined as \( \beta^* = 0.4765 \), reflecting the necessary compensation mechanism for biogas participation while maintaining household engagement. This result underscores the non-trivial interaction between economic incentives, investment thresholds, and the allocation of shared benefits within the REC structure.

\color{b}

\begin{figure}[H]
\centering
  \begin{subfigure}[b]{0.45\textwidth}
         \centering
         \includegraphics[width=\textwidth]{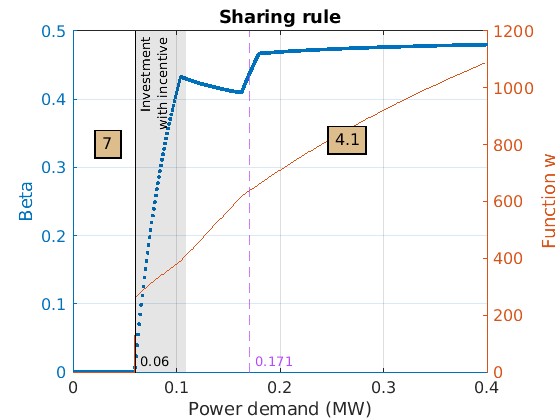}
         \caption{Optimal $\beta$ for different values of $d$}
         \label{ex_1_b_d}
     \end{subfigure}
    \hfill
     \begin{subfigure}[b]{0.45\textwidth}
         \centering
         \includegraphics[width=\textwidth]{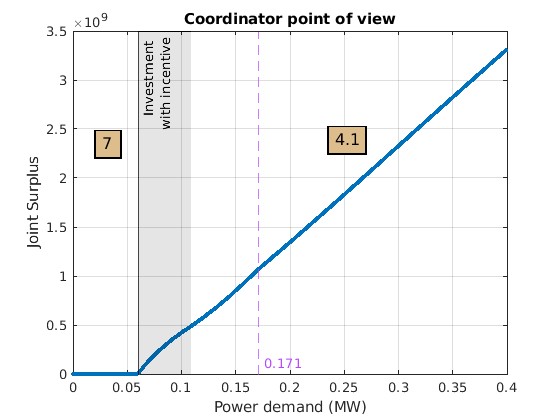}
         \caption{$F$ for different values of $d$}
         \label{ex_1_F_d}
     \end{subfigure}
        \caption{Sensitivity w.r.t. $d$}
        \label{b_F}
\end{figure}

\section{Conclusions}

The topic of renewable energy communities (REC) is expected to become of great relevance for the modern energy systems, since both self-production and self-consumption at a local scale reduce the problem of grid congestion on the large scale. Despite RECs being expected to be groups composed by a small number of
members, the problems that they are required to face and solve are quite sophisticated. 

In this paper we consider a REC composed of two specific types of members, namely a representative household and a biogas producer, which can be considered a typical configuration both in countryside villages as well as in urban peripheries. Moreover, we consider an advanced incentive model, 
namely the so-called "virtual framework" (adopted in Italy, with other countries being expected to follow its example), which introduces advantages, such as sustaining self-consumption and reducing energy poverty.

As we have shown, the viability of the REC, as well as the economic benefit accruing to its members, depends on the joint decision about the plants capacities and on the rule adopted to share the incentive among the members of the community.

A key finding of this analysis is that a centralized sharing rule is generally required for the two members, as splitting the incentive between them constitutes a zero-sum game, making a competitive equilibrium unattainable. In this study, the sharing rule is determined by a coordinator who ensures a balanced distribution of additional gains through a Nash bargaining function. Allowing the members to independently establish the sharing rule may undermine the viability and efficiency of the community, or even prevent its formation altogether.

Another important result is that the necessity of the incentive varies, depending on factors such as investment costs, the interest rate and the volatility of the stochastic variables influencing the members' profits. Its presence may be essential for one or both members, or entirely redundant.  From a policy perspective, these findings underscore the need for a carefully designed incentive framework that aligns financial support with actual economic necessity. Over-subsidization may lead to inefficient allocation of public funds and encourage speculative investments that do not contribute to long-term sustainability. Conversely, insufficient incentives could prevent the emergence of RECs, particularly in cases where investment in technologies such as biogas-to-power is only viable with financial support.  This study identifies the conditions under which the incentive serves as a genuine social benefit, rather than as a mere speculative opportunity.

A possible extension of this analysis involves incorporating the temporal variability of PV generation to distinguish between daylight and nighttime periods. Accounting for this variation would allow for a more realistic assessment of the integration between photovoltaic systems and more constant generation technologies, potentially enhancing the overall efficiency of the REC.

Additionally, another promising direction is the introduction of a dynamic decision-making framework for the biogas producer, enabling the flexible operation of the Gas-to-Power (G2P) turbine based on real-time electricity and gas price fluctuations. Allowing the biogas producer to strategically switch the turbine on or off depending on market conditions would not only improve its expected profitability but also potentially transform it into a net electricity consumer during periods when it exclusively sells gas. This shift could further strengthen the economic incentives for the biogas producer to participate in a REC, reinforcing the viability and sustainability of the community.


\color{black}

\appendix

\section{Appendix}

\subsection{Proofs}

\begin{proof}[Proof of Lemma \ref{lemmamin}]
First of all  we notice that $w$ depends on $y_b, y_h \geq 0$ only as parameters, as the nontrivial dependence is on $d$. For this reason, in this proof we will indicate explicitly only the dependence on $d$. Then if we assume that $\tau \sim Exp(\lambda)$ and is independent of $W_d$, then its randomness can be eliminated from the computation of $w$ in this way:
\begin{eqnarray*}
w(d) & = & \mathbb{E}\left[\int_{0}^{\tau} e^{-r s} \min \left\{ D^d(s), y_h + y_b \right\}ds \right] = \\
& = & \mathbb{E}\left[ \mathbb{E}\left[ \left. \int_{0}^{\tau} e^{-r s} \min \left\{ D^d(s), y_h + y_b \right\}ds \right| {\mathcal F}_\tau \right] \right] = \\
& = & \mathbb{E}\left[  \int_{0}^{\tau} e^{-r s} \mathbb{E}\left[ \left. \min \left\{ D^d(s), y_h + y_b \right\} \right| {\mathcal F}_\tau \right] ds  \right] = \\
& = & \mathbb{E}\left[  \int_{0}^{\tau} e^{-r s} \mathbb{E}\left[ \min \left\{ D^d(s), y_h + y_b \right\} \right] ds  \right] = \\
& = & \int_0^\infty \lambda e^{-\lambda u} \int_{0}^{u} e^{-r s} \mathbb{E}\left[ \min \left\{ D^d(s), y_h + y_b \right\} \right] ds\ du = \\
& = & \int_0^\infty   e^{-r s} \mathbb{E}\left[ \min \left\{ D^d(s), y_h + y_b \right\} \right]  \int_s^{\infty} \lambda e^{-\lambda u} du\ ds = \\
& = & \int_0^\infty   e^{-(r + \lambda) s} \mathbb{E}\left[ \min \left\{ D^d(s), y_h + y_b \right\} \right] ds = \\
& = & \mathbb{E}\left[\int_{0}^{\infty} e^{- (r + \lambda) s} \min \left\{ D^d(s), y_h + y_b \right\}ds \right]
\end{eqnarray*}
where we use the independence between $\tau$ and $W_d$ (thus $D$) in the fourth equality and the Tonelli theorem in the sixth and in the eighth. 

We can easily check that, with this new formulation, $w$ satisfies the ordinary differential equation (ODE)
\begin{equation} \label{ODE}
(r + \lambda) w(d) - \mathcal{L} w(d) = \min(d,y_h + y_b) 
\end{equation}
where $\mathcal{L} u(d) = \left( \mu_d + \frac{\sigma_d^2}{2} \right) d u'(d) + \frac{\sigma_d^2}{2} d^2 u''(d)$. In fact, 
the general solution of Equation \eqref{ODE} is 
\begin{gather}
w(d) = \begin{cases}
Ad^{m_1} + Bd^{m_2} + \frac{d}{r_d + \lambda} & d < y_h + y_b\\
Cd^{m_1} + Dd^{m_2} + \frac{y_h +y_b}{r + \lambda} & d \geq y_h +y_b
\end{cases}
\end{gather}
where $A$, $B$, $C$ and $D$ are constants to be determined.
As we want Equation \eqref{w(d)} to hold, we want $w(0) = 0$,  therefore we must have $A = 0$. On the other hand, it is also clear from Equation \eqref{w(d)} that $w(y_h,y_b,d) \leq \frac{y_h + y_b}{r}$, thus $w$ must stay bounded when $d \rightarrow \infty$, which gives $D=0$. Since Equation \eqref{ODE} is a linear second-order ODE with continuous coefficients, we expect its solution to be of class $C^2({\mathbb R}^+)$. For this reason, we impose the value matching and smooth pasting conditions at
at $d = y_h + y_b$, which imply
\begin{gather} \label{BC}
\begin{cases}
B (y_h +y_b)^{m_2} + \frac{y_h +y_b}{r_d + \lambda} = C (y_h +y_b)^{m_1} + \frac{y_h +y_b}{r + \lambda}\\
B m_2 (y_h +y_b)^{m_2-1} + \frac{1}{r_d + \lambda} = C m_1 (y_h +y_b)^{m_1-1} ,
\end{cases}
\end{gather}
which gives
\begin{gather}
\begin{cases}
B = (y_h + y_b)^{1-m_2} B_1\\
C = (y_h + y_b)^{1-m_1}C_1,
\end{cases}    
\end{gather}
with $B_1$ and $C_1$ as in Equation \eqref{B1C1}. With this determination, one can easily check that the function $w$ is increasing and concave on ${\mathcal R}^+$, and $C^2$ in $d$. We can easily check that $w$ is $C^2$ also in $y_h$ and $y_b$. In fact, it is outside of the region $\{y_b + y_h = d\}$. Moreover,
\begin{equation} \label{wc1}
\left. w_{y_b} \right|_{y_b + y_h = d^+} = 
\left. w_{y_h} \right|_{y_b + y_h = d^+} = B_1 (1 - m_2) = \left. w_{y_b} \right|_{y_b + y_h = d^-} = \left. w_{y_h} \right|_{y_b + y_h = d^-} = C_1 (1 - m_1) + \frac1{r + \lambda} 
\end{equation}
which is easily obtained from the system \eqref{BC} by simplifying for $(y_b + y_h)$ in the first equation and subtracting the second. The fact that the second derivatives are also continuous at $\{y_b + y_h = d\}$ can be verified with some algebra. 

By applying the Ito formula to the process $(e^{- (r + \lambda) t} w(D(t)))_t$ from $0$ to $T$, we have 
\begin{eqnarray*}
\lefteqn{ e^{- (r + \lambda) T} w(D^d(T)) = w(d) +\int_{0}^T e^{- (r + \lambda) s} \left( \mathcal{L} w(D^d(s)) - (r + \lambda) w(D^d(s)) \right)ds } \\
&   & + \int_0^T  \sigma_d e^{- (r + \lambda) s} D^d(s) w'(D^d(s))dW_d(s) \\
& = & w(d) - \int_{0}^T e^{- (r + \lambda) s}  \min \{ D^d(s), y_h + y_b \} ds + \int_0^T  \sigma_d e^{- (r + \lambda) s} D^d(s) w'(D^d(s))dW_d(s).
\end{eqnarray*}
One can easily check that $w$ is concave, thus $|w'(d)| \leq w'(0) < + \infty$ for all $d \geq 0$: this implies 
\begin{eqnarray*}
\mathbb{E}\left[\int_{0}^{T} e^{-2(r + \lambda) s} D^2(t) (w'(D(t))^2 dt \right] & \leq & \mathbb{E}\left[\int_{0}^{T} e^{-2(r + \lambda) s} D^2(t) (w'(0))^2 dt \right]\\
& = & \frac{1}{r_d^2} \int_{0}^{T} e^{- 2s(r + \lambda - \mu_d)} \mathbb{E} \left[ e^{2 \sigma_d W(s)} \right]ds\\
& = & \frac{1}{r_d^2} \int_{0}^{T} e^{- 2s(r + \lambda - \mu_d - \sigma_d^2)} ds < \infty
\end{eqnarray*}
thus the expectation of the stochastic integral above is zero  and we have that, for all $T > 0$, 
\begin{equation} \label{prelimit}
w(d) = \mathbb{E}\left[e^{- (r + \lambda) T} w(D^d(T)) \right] + \mathbb{E}\left[ \int_{0}^T e^{- (r + \lambda) s}  \min \{ D^d(s), y_h + y_b \} ds \right] 
\end{equation}
Finally, since for all $d \geq 0$ we have
\begin{eqnarray*}
0 \leq w(d) \leq \lim_{d \to + \infty} w(d) = \frac{y_h + y_b}{r + \lambda}, 
\end{eqnarray*}
then
\begin{eqnarray*}
\lim_{T \rightarrow \infty } \left| \mathbb{E}\left[ e^{- (r + \lambda) T} w(D^d(T))\right] \right| \leq \lim_{T \rightarrow \infty } e^{- (r + \lambda) T} \frac{y_h + y_b}{r + \lambda} = 0,
\end{eqnarray*}
and by letting $T \to + \infty$ in Equation \eqref{prelimit}, we get the result in Equation \eqref{w(d)}. 
\end{proof}

\vspace{2ex}

\begin{proof}[Proof of Proposition \ref{proph}]
We can find the point of maximum of $J_h$ using a first order condition. The partial derivative of $J_h$ w.r.t. $y_h$ is given by

\begin{gather} \label{dJh}
    \frac{\partial J_h}{\partial y_h} = \begin{cases}
    g_h + Z \beta B_1 d^{m_2}(1 - m_2)(y_h + y_b)^{-m_2} & d < y_h + y_b \\
    g_h + Z \beta ( C_1 d^{m_1}(1 - m_1)(y_h + y_b)^{-m_1} +  \frac{1}{r + \lambda}) & d \geq y_h + y_b.
    \end{cases}
\end{gather}

\noindent $\frac{\partial J_h}{\partial y_h}$ is continuous in $y_h$, then we search for the critical point $\frac{\partial J_h}{\partial y_h} = 0$. As $y_h \in [0, \theta_h]$ we also have to consider the extreme points $0$ and $\theta_h$. 

Since $J_h$ is $C^1$ and concave in $y_h$, we can determine whether the maximum is attained when the aggregate installation is greater than the demand or not, by observing the sign of $\frac{\partial J_h}{\partial y_h} |_{y_h = d - y_b}$. In fact $\frac{\partial J_h}{\partial y_h} |_{y_h = d - y_b} \lesseqgtr 0$  if and only if the maximum $y_h^* \lesseqgtr d - y_b$. 

We have that 
$$ \left. \frac{\partial J_h}{\partial y_h} \right|_{y_h = d - y_b} = g_h + Z \beta B_1 (1 - m_2) $$
thus, since $B_1 < 0$ and $m_2 > 1$, we have that $\frac{\partial J_h}{\partial y_h} |_{y_h = d - y_b} > 0$ if and only if $\beta > \beta_h$, this latter being defined in Equation \eqref{betah}. 
%
%
%
%

Let us move into the solution of problem \eqref{maxjh}. Let us suppose $\beta > \beta_h$, so that $\frac{\partial J_h}{\partial y_h} |_{y_h = d - y_b} > 0$: then we know that the maximum is attained when the installation exceeds the demand, i.e. when $d < y_b + y_h$.  The first-order condition \eqref{dJh} are solved, in the case $d < y_h + y_b$, for
$$  Z \beta B_1 d^{m_2}(m_2 - 1)(y_h + y_b)^{-m_2}  =  g_h $$
As the left-hand side is always negative, the first-order condition has a solution only if $g_h < 0$. In this case, $\beta > \beta_h > 0$ and the solution is
$$ y_h^* = d\left( \frac{Z\beta B_1 (m_2 - 1) }{g_h }\right)^{1/m_2} - y_b { = d\left( \frac{\beta}{\beta_h}\right)^{1/m_2} - y_b}. $$
\noindent If $y_h^* \in [0,  \theta_h ]$, then $y_h^*$ is the maximum. { If $y_h^* < 0$ then the maximum is 0 and if $y_h^* > \theta_h$ then the maximum }. is $\theta_h$. Moreover, if $ ZB_1( m_2 - 1) < g_h $, then $\beta_h < 1$ 
. {Summarizing, 

\begin{eqnarray}
    y_h^* = \max \left\{ \min \left\{  d\left( \frac{\beta}{\beta_h}\right)^{1/m_2} - y_b , \theta_h \right\} , 0 \right\}.
\end{eqnarray}
}

\noindent Instead, if $g_h \geq 0$, then the first-order condition is not satisfied, $\frac{\partial J_h}{\partial y_h}$ is always positive and the maximum of $J_h$ is reached for $ y_h^* = \theta_h$. In this case $\beta_h < 0$. 

Let us now suppose that $\beta \leq \beta_h$, i.e. $\frac{\partial J_h}{\partial y_h} |_{y_h = d - y_b} \leq 0$. Then for $d \geq y_h + y_b$, the first order condition is solved for

$$Z\beta  C_1 d^{m_1} (m_1 - 1) (y_h + y_b)^{-m_1}  = g_h + \frac{Z \beta}{r + \lambda},$$

\noindent as the left hand side is always positive, the first order condition has a solution only if  $g_h + \frac{ Z \beta}{r + \lambda}  > 0$, and the solution is
$$ y_h^* = d\left( \frac{Z\beta C_1 (m_1 - 1) }{g_h + \frac{Z\beta}{r + \lambda} }\right)^{1/m_1} - y_b { = d\left(  \frac{ \frac{g_h}{\beta_h}   + \frac{Z}{r + \lambda}  }{\frac{g_h}{\beta}  + \frac{Z}{r + \lambda} }\right)^{1/m_1} - y_b}. $$
\noindent moreover if $g_h < 0$, then $\beta_h >0$. If $y_h^* \in [0, \theta_h]$, then { $y_h^*$} is the maximum. If $y_h^* < 0$ then the maximum will be attained at zero, and if $y_h^* > \theta_h$ then the maximum will be attained at $\theta_h$. 
%
%
{ Summarizing, 

\begin{eqnarray}
    y_h^* = \max \left\{ \min \left\{  d\left(  \frac{ \frac{g_h}{\beta_h}   + \frac{Z}{r + \lambda}  }{\frac{g_h}{\beta}  + \frac{Z}{r + \lambda} }\right)^{1/m_1} - y_b , \theta_h \right\} , 0 \right\}.
\end{eqnarray}
}
On the other hand, if $g_h + \frac{ Z \beta}{r + \lambda} \leq 0$, the derivative is always negative and the maximum is { always} attained at $y_h^* = 0$. Again in this case $\beta_h > 0$. 
\end{proof}

\vspace{2ex}

\begin{proof}[Proof of Proposition \ref{propb}]
We can find the maximum point of $J_b$ using a first order condition. The partial derivative of $J_b$ w.r.t. $y_b$ is given by

\begin{gather} \label{dJb}
    \frac{\partial J_b}{\partial y_b } = \begin{cases}  
    g_b + Z (1 -\beta) \left( B_1(1-m_2)(y_h + y_b)^{-m_2} d^{m_2} \right) & d < y_h + y_b\\
   g_b + Z (1 -\beta) \left( C_1 (1-m_1)(y_h + y_b)^{-m_1}d^{m_1} +  \frac{1}{r + \lambda} \right) & d \geq y_h + y_b.
    \end{cases}
\end{gather}

\noindent $\frac{\partial J_b}{\partial y_b}$ is continuous in $y_b$, then we search for the critical point $\frac{\partial J_b}{\partial y_b} = 0$. As $y_b \in [0, \theta_b]$ we also have to consider the extreme points $0$ and $\theta_b$. 

Since $J_b$ is $C^1$ in $y_b$ and it is concave, we can determine whether the maximum is attained when the aggregate installation exceeds the demand or not, by observing the sign of $\frac{\partial J_b}{\partial y_b} |_{y_b = d - y_h}$. In fact $\frac{\partial J_b}{\partial y_b} |_{y_b = d - y_h} \lesseqgtr 0$  if and only if the maximum $y_b^* \lesseqgtr d - y_h$. 

{
We have that
$$ \left. \frac{\partial J_b}{\partial y_b} \right|_{y_b = d - y_h} = g_b + Z (1 -\beta)  B_1(1-m_2) = 
   g_b + Z (1 -\beta) \left( C_1 (1-m_1) +  \frac{1}{r + \lambda} \right) $$
thus, since $B_1 < 0$ and $m_2 > 1$, we have that $\frac{\partial J_b}{\partial y_b} |_{y_b = d - y_h} > 0$ if and only if $1 - \beta > 1 - \beta_b$ with $\beta_b$ defined in Equation \eqref{betab}.
}
%
%


Let us move into the solution of problem \eqref{maxjb}. Let us suppose $\beta < \beta_b$, i.e. $\frac{\partial J_b}{\partial y_b} |_{y_b = d - y_h} > 0$, then we know that the maximum is attained when the installation exceeds  the demand, i.e. 
$d < y_h + y_b$: the first-order condition is then solved for 
$$ g_b =  Z (1 -\beta) \left( B_1(m_2-1)(y_h + y_b)^{-m_2} d^{m_2} \right). $$

\noindent Since the right-hand side is always negative, the first order condition has a solution only if $g_b < 0$ and the solution is 
\begin{eqnarray*}
y_b^* = d \left( \frac{Z(1 - \beta) B_1  (m_2 - 1) }{g_b } \right)^{1/m_2}  - y_h { = d \left( \frac{1 - \beta}{1 - \beta_b } \right)^{1/m_2}  - y_h}.
\end{eqnarray*}
If $y_b^* \in [0, \theta_b]$, then it is the maximum. If $y_b^* > \theta_b$, then we select $\theta_b$ as optimal point, instead if $y_b^* < 0$, we select $y_b^* = 0$. Moreover, if $g_b > Z B_1 (m_2 - 1)$, then $\beta^b > 0$. {Summarizing, 

\begin{eqnarray}
    y_b^* = \max \left\{ \min \left\{  d \left( \frac{1 - \beta}{1 - \beta_b } \right)^{1/m_2} - y_h , \theta_b \right\} , 0 \right\}.
\end{eqnarray}
}

\noindent In the case $g_b \geq 0$, the derivative is always positive and the maximum is attained at $y_b^* = \theta_b$. 

In analogy with the previous case, let us now suppose $\beta \geq \beta_b$, i.e. $\frac{\partial J_b}{\partial y_b} |_{y_b = d - y_h} \leq 0$, in this case $d \geq y_h + y_b$. 
The first order condition is now solved for 
$$ g_b + \frac{ Z (1 -\beta) }{r + \lambda} =  Z (1 -\beta) \left( C_1(m_1-1)(y_h + y_b)^{-m_1} d^{m_1}  \right).$$
Since the right-hand side is always positive, the first order condition is solved only if $g_b + \frac{ Z (1 -\beta) }{r + \lambda} > 0$ and the solution is
\begin{eqnarray}
y_b^* = d \left( \frac{Z(1 - \beta)C_1 (m_1 - 1) }{ g_b + \frac{Z(1-\beta)}{r+ \lambda} } \right)^{1/m_1} - y_h  = d \left( \frac{\frac{g_b}{1- \beta_b} + \frac{Z}{r+ \lambda}}{ \frac{g_b}{1- \beta} + \frac{Z}{r+ \lambda}}  \right)^{1/m_1} - y_h
\end{eqnarray}
If $y_b^* \in [0, \theta_b]$, then it is the maximum. If $y_b^* > \theta_b$, then we select $\theta_b$ as optimal point, instead if $y_b^* < 0$, we select $y_b^* = 0$. If, moreover, $g_b < 0$, then $\beta^b < 1$. Furthermore, if it is verified that $Z B_1 (m_2 - 1) < g_b $, then $\beta^b > 0$. {Summarizing, 

\begin{eqnarray}
    y_b^* = \max \left\{ \min \left\{  d \left( \frac{\frac{g_b}{1- \beta_b} + \frac{Z}{r+ \lambda}}{ \frac{g_b}{1- \beta} + \frac{Z}{r+ \lambda}} \right)^{1/m_1} - y_h , \theta_b \right\} , 0 \right\}.
\end{eqnarray}
}

In the case $g_b + \frac{Z(1-\beta)}{r+ \lambda}\leq 0$, the derivative is always negative and the maximum is attained at $y_b^* = 0$. 
\end{proof}

\vspace{2ex}

\begin{proof}[Proof of Proposition \ref{ne}] 
{ We start by proving that the equilibria presented in \eqref{NYC11},\eqref{NYC12},\eqref{NYC13}, \eqref{NYC14}, \eqref{NYC15}, \eqref{NYC16}, \eqref{NYC17}, \eqref{NYC18}, \eqref{NYC2} and \eqref{NYC3} are Nash equilibria. In all these cases the community is performed if the biogas investment is positive. {Moreover, notice that, in all the cases in equations \eqref{NYC11},\eqref{NYC12},\eqref{NYC13}, \eqref{NYC14}, \eqref{NYC15}, \eqref{NYC16}, \eqref{NYC17} and \eqref{NYC18}, one of the two players ($b$ or $h$) has an optimal installation level which does not depend on the other player's decision, as either its net unit profit $g_{b,h}$ is strictly positive, thus the corresponding optimal strategy $y_{b,h}^* = \theta_{b,h}$, or its marginal cost is too low to install anything even with incentives, thus $y_{b,h}^* = 0$. In all these cases, the presented Nash equilibria follow quite easily. }


{We now pass to the only nontrivial case, i.e. when 
$g_h + \frac{Z \beta}{r + \lambda} > 0 > g_h$ {\color{red} and} $g_b + \frac{Z(1 - \beta)}{r + \lambda} > 0 > g_b$.
First of all, in this case we have $\beta_h > 0$ and $1 - \beta_b > 0$, thus one can easily verify from Equation \eqref{betan} that $\beta_n > \beta_h$ if and only if $\beta_h < \beta_b$, and $\beta_n < \beta_b$ if and only if $\beta_h < \beta_b$, hence $\beta_n$ must lie in the interval with extrema $\beta_b$ and $\beta_h$. Moreover, we see that 
the value of $\beta_n$ comes from the computation of the best responses of the household and biogas producer. From Propositions \ref{besthousehold} and \ref{bestbiogas} we know that, if $\beta_h < \beta_b$, if we take any $\beta \in (\beta_h,\beta_b)$, 
the best responses are 
\begin{eqnarray} \label{auxNa}
   y_h^* & = & \max \left\{ \min \left\{ \displaystyle d\left( \frac{\beta}{\beta_h}\right)^{1/m_2} - y_b , \theta_h \right\} , 0 \right\}, \\
   y_b^* & = &  \max \left\{ \min \left\{  d \left( \frac{1 - \beta}{1 - \beta_b } \right)^{1/m_2} - y_h , \theta_b \right\} , 0 \right\}.
\end{eqnarray}
Assume now that we have an internal Nash equilibrium, i.e.
\begin{eqnarray}
   y_h^* & = &  \displaystyle d\left( \frac{\beta}{\beta_h}\right)^{1/m_2} - y_b^* ,  \label{compbn} \\
   y_b^* & = &   d \left( \frac{1 - \beta}{1 - \beta_b } \right)^{1/m_2} - y_h^* .  \label{compbn2}
\end{eqnarray}
\noindent The system  given by Equations \eqref{compbn} and \eqref{compbn2} 
has solutions if and only if 
\begin{eqnarray*}
    y_h^* + y_b^* = d \left( \frac{\beta}{\beta_h}\right)^{1/m_2} =  d \left( \frac{1 - \beta}{1 - \beta_b } \right)^{1/m_2},
\end{eqnarray*}
\noindent which, after some algebra, is true if and only if  $\beta = \beta_n$ as defined in Equation \eqref{betan}. 
Conversely, in the case $\beta_h < \beta_b$, the same argument brings us to 
$$     y_h^* + y_b^* = d\left( \displaystyle \frac{  \frac{g_h}{\beta_h}   + \frac{Z}{r+ \lambda}  }{\frac{g_h}{\beta}  + \frac{Z}{r+ \lambda} }\right)^{1/m_1} 
= d\left( \displaystyle \frac{  \frac{g_b}{1-\beta_b}   + \frac{Z}{r+ \lambda}  }{\frac{g_b}{1-\beta}  + \frac{Z}{r+ \lambda} }\right)^{1/m_1}, $$
However, by Remark 3.10, we have $\frac{g_h}{\beta_h} = \frac{g_b}{1-\beta_b}$,, which after some algebra brings, as in the previous case, a solution if and only if $\beta = \beta_n$ as defined in Equation \eqref{betan}. }

{To fix the ideas, suppose now $\beta_n \in (\beta_h, \beta_b)$ (the same arguments can be applied for $ \beta_n \in (\beta_b, \beta_h)$). Assume also for simplicity that $d \left(\frac{1-\beta_n}{1-\beta_b} \right)^{1/m_2}$ is not greater than both $\theta_b$ and $\theta_h$. In this case we check if any strategy of the form
\begin{eqnarray*}
    y_h^* & = & \alpha d \left( 
    \frac{1-\beta_n}{1-\beta_b} \right)^{1/m_2} , \\ 
    y_b^* & = & (1 - \alpha) d \left( \frac{1-\beta_n}{1-\beta_b} \right)^{1/m_2},
\end{eqnarray*}
\noindent with $\alpha \in (0,1)$, is a Nash equilibrium by checking Equation \eqref{NasDef}. 
For the household profit we have 
\begin{eqnarray} \label{NEb}
    J_h(x_v, x_c , p ,d , y_h, y_b^*, \beta) & = &  y_h g_h - \frac{x_c d}{r_{cd}} + \beta_n Z w(y_h, y_b^* , d)\noindent 
\end{eqnarray}
which we can maximize in $y_h$ by using Proposition 3.5, finding that 
\begin{eqnarray}\label{NEHb}
    \argmax_{y_h \in [0, \theta_h]} J_h(x_v, x_c , p ,d , y_h, y_b^*) & = & \alpha d\left( \frac{1-\beta_n}{1-\beta_b} \right)^{1/m_2}.
\end{eqnarray}
\noindent On the other hand, for the biogas profit we have
\begin{eqnarray*}
    J_b(x_v, x_c , p ,d , y_h^*, y_b) & = &  y_b g_b + \frac{p b K_g}{r_p} + (1 - \beta_n) Z w(y_h^*, y_b , d)
\end{eqnarray*}
which we can maximize in $y_b$ by using Proposition 3.7, finding that 
\begin{eqnarray} \label{NEBb}
    \argmax_{y_b \in [0, \theta_b]} J_b(x_v, x_c , p ,d , y_h^*, y_b) & = & (1 - \alpha) d\left( \frac{1-\beta_n}{1-\beta_b} \right)^{1/m_2}.
\end{eqnarray}
\noindent By \eqref{NEHb} and \eqref{NEBb} the strategies \eqref{NEb} satisfy Equation \eqref{NasDef} and therefore any linear combination $\alpha \in (0,1)$ is a Nash equilibria.
}

\noindent In all the other possible cases in \eqref{auxNa}, the system is solvable for any $\beta \in (0,1)$  and the value for the optimum $\beta$ is computed by solving the coordinators problem \eqref{cooprob}.

    }
\end{proof}

\subsection{Parameter estimation}
\begin{table}[H]
\begin{small}
    \centering
    \begin{tabular}{|c|r|r|c|r|r|c|r|r|}
    \hline
       \multicolumn{3}{|c|}{Demand}   & \multicolumn{3}{|c|}{Electricity price} & \multicolumn{3}{|c|}{Gas price}  \\
        \hline
        Freq. (1/h)  & \multicolumn{2}{|c|}{Amplitude}  & Freq. (1/h) & \multicolumn{2}{|c|}{Amplitude}  & Freq. (1/d) & \multicolumn{2}{|c|}{Amplitude}  \\
        \hline
        & $A$ & $B$ & &  $A$ & $B$ &  & $A$ & $B$  \\
        \hline
        0.00023 &  2575.5 & 583.2 & 1.12e-05 & -0.47258 & 0.33134 & 0.0003817 & -0.36363 & 0.46082  \\
        0.00034 & 1285.6 & -630.7 & 2.25e-05 & -0.52073 & -0.06719 & 0.0007634 & -0.40630 & -0.56797\\
        0.00057 & 0 & -1391.9 & 3.37e-05 &  0.06854 & -0.32286 & 0.001145 & 0.16201 & -0.40630  \\
        0.00068 & 836.4& 458.5 & 6.737e-05 & -0.05996 & 0.06290 &  &  & \\
        0.00092 & -314.4 & 0 & 7.86e-05 & -0.04909 & 0.06341 &  &  & \\
        0.00594  & -459.8 & 666.4 & 4.49e-05 & 0.09773 & 0  &   &  &  \\
        0.00595 & -2803.1 &  2344.5 & 0.0001123 & -0.16543 & 0 &  &  & \\
        0.01191 & -1344.8 & -747.5 & 0.00022456 & 0 & 0.05539 &  &  & \\
        0.01786 & 344.5& 0 & 0.00029192 & 0 & 0.05420  &  & &  \\
        0.02976 & 523.1 & 930.8 & 0.0059508  & -0.05448 & 0.07790 &  &  &   \\
        0.03571 & 434.4 & -992.2 & 0.0119016 & 0.04609 & 0 &  &  &  \\
        0.04155 & -485.4 & 0 & 0.04155439 & -0.04806 & 0 &  &  & \\
        0.04167 & -5062.2 & 1014.9 & 0.0416667 & -0.10614  & 0 &  &  &\\
        0.04762 & 0 & -893.2 &  0.0833333 & -0.14971  & -0.04988  &  &  &  \\
        0.08333 & -1583.1 & -2257.3 & 0.125 &0.03693 & 0 &  &  &  \\
        0.125 & 650.0  & 0 &  & &  &  &  & \\
        \hline
    \end{tabular}
    \caption{Significant frequencies }
    \label{fd}
    \end{small}
\end{table}





\begin{thebibliography}{99}

\bibitem{VEC} Abada, I., Ehrenmann, A., Lambin, X. (2017). On the viability of energy communities. \textit{Energy Policy Research Group, University of Cambridge}.

\bibitem{P2PStak} Anoh, K., Maharjan, S., Ikpehai, A., Zhang, Y., Adebisi, B. (2020).
Energy Peer-to-Peer Trading in Virtual Microgrids in Smart Grids: A Game-Theoretic Approach.
\textit{IEEE Transactions on Smart Grid},
\textbf{11(2)} 1264-1275.

\bibitem{BBCMV} Bergemann, D., Bertolini, M., Castellini, M., Moretto, M.,
 Vergalli, S. (2022). Renewable energy communities, digitalization and
information. Nota di Lavoro 037.2022, Milano, Italy: Fondazione Eni Enrico
Mattei. Available at \verb|https://papers.ssrn.com/sol3/papers.cfm?abstract_id=4294765|

\bibitem{ts}
 Bloomfield, P.
 (2007).
 \textit{Fourier Analysis of Time Series: An Introduction.}
 Wiley series in probability and mathematical statistics.

\bibitem{brigo}
  Brigo, D., Dalessandro, A., Neugebauer, M., Triki, F. (2008). A Stochastic Processes Toolkit for Risk Management.  Available at \url{https://arxiv.org/abs/0812.4210}.

\bibitem{RI} Candelise, C., Ruggieri, G.
(2020). 
Status and Evolution of the Community Energy Sector in Italy.
\textit{Energies},
\textbf{13 (8)} 1888.

\bibitem{ECIt1} Cutore, E., Volpe, R.,  Sgroi, R., Fichera, A.
(2023).
Energy management and sustainability assessment of renewable energy communities: The Italian context.
\textit{Energy Conversion and Management},
\textbf{278} 116713.

\bibitem{ECIt3} {De Juan-Vela}, P., Alic, A., Trovato, V.
(2023).
Monitoring the Italian transposition of the EU regulation concerning renewable energy communities and the relevant policies for battery storage.
\textit{Journal of Cleaner Production},
\textbf{425} 138937.

\bibitem{ECIt2} Dimovski, A., Moncecchi, M., Merlo, M.
(2023).
Impact of energy communities on the distribution network: An Italian case study.
\textit{Sustainable Energy, Grids and Networks},
\textbf{35} 101148.

\bibitem{2018/2001} European Parliament. Directive (EU) 2018/2001 of the European Parliament and of the Council of 11 December 2018 on the promotion of the use of energy from renewable sources. Available at \url{https://eur-lex.europa.eu/eli/dir/2018/2001/oj}
%
\bibitem{2019/944} European Parliament. Directive (EU) 2019/944 of the European Parliament and of the Council of 5 June 2019 on common rules for the internal market for electricity and amending Directive 2012/27/EU. Available at 
\url{https://eur-lex.europa.eu/eli/dir/2019/944/oj}

\bibitem{CGEC} Feng, C., Wen, F., You, S., Li, Z., Shahnia, F., Shahidehpour, M.
(2020).
Coalitional Game-Based Transactive Energy Management in Local Energy Communities.
\textit{IEEE Transactions on Power Systems}, 
\textbf{35(3)} 1729-1740.

\bibitem{RevEC} Gjorgievski, V.Z., Cundeva, S., Georghiou, G.E.
(2021).
Social arrangements, technical designs and impacts of energy communities: A review.
\textit{Renewable Energy},
\textbf{169} 1138-1156.

\bibitem{IT1} Italian Government. Law decree 162/2019. Urgent provisions on the extension of legislative terms, organization of the public administrations and technological innovation.
 \url{https://www.gazzettaufficiale.it/eli/id/2019/12/31/19G00171/sg}

\bibitem{IT2} Italian Parliament. Law 8/2020, Conversion to Law of the Law decree 162/2019, with amendments. 
\url{ https://www.gazzettaufficiale.it/eli/id/2020/02/29/20G00021/sg}

\bibitem{IT3} Italian Ministry of Ecological Transition (MITE). Decree 2020, September 16. On the identification of the incentive tariff for the remuneration of renewable energy plants.
 \url{ https://www.gazzettaufficiale.it/eli/id/2020/11/16/20A06224/sg}

\bibitem{IT4} Italian Ministry of Environment and Energy Safety. Decree n. 414, 2023 December 7, Identification of an incentive tariff for renewable energy plants included in renewable energy communities. \url{ https://www.mase.gov.it/sites/default/files/Decreto%20CER.pdf}

\bibitem{CSG2} Li, L.
(2020).
Optimal Coordination Strategies for Load Service Entity and Community Energy Systems Based on Centralized and Decentralized Approaches.
\textit{Energies},
\textbf{13 (12)} 3202.

\bibitem{GTMLEC} Lilliu, F., Denysiuk, R., Reforgiato Recupero, D., Vinyals, M.
(2021).
A Game-Theoretical Incentive Mechanism for Local Energy Communities.
\textit{Agents and Artificial Intelligence},
Springer International Publishing, 52-72.

\bibitem{RECbook} L\"obbe, S., Sioshansi, F., Robinson, D. (2022). \textit{Energy Communities: Customer-Centered, Market-Driven, Welfare-Enhancing?}
Elsevier Science.

\bibitem{VAB} Manuel De Villena, M., Aittahar, S., Mathieu, S., Boukas, I., Vermeulen, E., Ernst, D. (2022). Financial Optimization of Renewable Energy Communities Through Optimal Allocation of Locally Generated Electricity. 
\textit{IEEE Access} \textbf{10} 77571-77586.

\bibitem{MDCP} Mitridati, L., Kazempour, J., Pinson, P. (2021).
Design and game-Theoretic analysis of community-Based market mechanisms in heat and electricity systems. \textit{Omega}, \textbf{99} 102 - 177.

\bibitem{MMM} Moncecchi, M., Meneghello, S., Merlo, M. 
(2020). 
A Game Theoretic Approach for Energy Sharing in the Italian Renewable Energy Communities. 
\textit{Applied Sciences}, \textbf{10(22)}, 8166.

\bibitem{ECIt4} Musolino, M., Maggio, G., D'Aleo, E., Nicita, A.
(2023).
Three case studies to explore relevant features of emerging renewable energy communities in Italy.
\textit{Renewable Energy},
\textbf{210} 540-555.

\bibitem{MCA} Norbu, S., Couraud, B.,  Robu, V., Andoni, M., Flynn, D.
(2021).
Modelling the redistribution of benefits from joint investments in community energy projects.
\textit{Applied Energy},
\textbf{287} 116575.

\bibitem{P2Pstak2} Paudel, A., Chaudhari, K., Long, C., Gooi, H. B.
(2019). 
Peer-to-Peer Energy Trading in a Prosumer-Based Community Microgrid: A Game-Theoretic Model. \textit{ IEEE Transactions on Industrial Electronics}
\textbf{66(8)}  6087-6097.

\bibitem{hans}
 Peters, H.
 (2008).
 \textit{Game Theory: A multilevel approach}. 
 Springer Berlin, Heidelberg.

\bibitem{pham}
  Pham, H.
  (2009).
  \textit{Continuous-time Stochastic Control and Optimization with Financial Applications}.
  Springer, Berlin, Heidelberg.
  
\bibitem{pirjol}
  Pirjol, D., Zhu, L. (2016). Discrete sums of geometric Brownian motions, annuities and Asian options. \textit{Insurance: Mathematics and Economics}, \textbf{70} 19-37.
  
\bibitem{CGS} Safdarian, A., Astero, P., Baranauskas, M., Keski-Koukkari, A., Kulmala, A. (2021). Coalitional Game Theory Based Value Sharing in Energy Communities. \textit{IEEE Access} \textbf{9} 78266-78275.

\bibitem{BCCGEC} Toderean, L., Chifu, V. R., Cioara, T., Anghel, I., Pop, C. B.
(2023).
Cooperative Games Over Blockchain and Smart Contracts for Self-Sufficient Energy Communities.
\textit{IEEE Access}, 
\textbf{11} 73982-73999.

\bibitem{P2PGame} Tushar, W., Saha, T. K., Yuen, C., Morstyn, T., McCulloch, M.D., Poor, H.V., Wood, K.L.
(2019).
 A motivational game-theoretic approach for peer-to-peer energy trading in the smart grid.
\textit{Applied Energy},
\textbf{243} 10-20.

\bibitem{WCRS} Tushar, W., Yuen, C., Mohsenian-Rad, H., Saha, T., Poor, H. V., Wood, K. L. (2018). Transforming Energy Networks via Peer-to-Peer Energy Trading: The Potential of Game-Theoretic Approaches.
\textit{IEEE Signal Processing Magazine}, \textbf{35(4)}  90-111.

\bibitem{EIA} U.S. Energy Information Administration. International Energy Outlook 2021. Available online at \url{ https://www.eia.gov/outlooks/ieo/pdf/IEO2021\textunderscore ChartLibrary\textunderscore full.pdf}

\bibitem{ECIt} Zatti, M., Moncecchi, M., Gabba, M., Chiesa, A., Bovera, F., Merlo, M.
(2021).
Energy Communities Design Optimization in the Italian Framework.
\textit{Applied Sciences},
\textbf{11(11)} 5218.

\bibitem{GastEle} https://www.eia.gov/energyexplained/units-and-calculators/energy-conversion-calculators.php

\end{thebibliography}
\end{document}